\documentclass[preprint2,times,tighten]{aastex6}
\usepackage{amsmath,amstext}
\usepackage[all]{hypcap} 

\usepackage{bm}
\usepackage{subfigure}
\usepackage{multirow}
\usepackage{etoolbox}
\usepackage{lineno}
\usepackage{units}
\usepackage{xspace}
\usepackage{float}

\shorttitle{Observation and Characterization of a Cosmic Muon Neutrino Flux}
\shortauthors{M. G. Aartsen et al.}

\begin{document}

\journalinfo{Version 3.0, dated \today}
\submitted{}

\title{Observation and Characterization of a Cosmic Muon Neutrino Flux from the Northern Hemisphere using six years of IceCube data}

\author{
IceCube Collaboration:
M.~G.~Aartsen\altaffilmark{1},
K.~Abraham\altaffilmark{2},
M.~Ackermann\altaffilmark{3},
J.~Adams\altaffilmark{4},
J.~A.~Aguilar\altaffilmark{5},
M.~Ahlers\altaffilmark{6},
M.~Ahrens\altaffilmark{7},
D.~Altmann\altaffilmark{8},
K.~Andeen\altaffilmark{9},
T.~Anderson\altaffilmark{10}
% ...
}
\altaffiltext{1}{Department of Physics, University of Adelaide, Adelaide, 5005, Australia}
\altaffiltext{2}{Physik-department, Technische Universit\"at M\"unchen, D-85748 Garching, Germany}
\altaffiltext{3}{DESY, D-15735 Zeuthen, Germany}
\altaffiltext{4}{Dept.~of Physics and Astronomy, University of Canterbury, Private Bag 4800, Christchurch, New Zealand}
\altaffiltext{5}{Universit\'e Libre de Bruxelles, Science Faculty CP230, B-1050 Brussels, Belgium}
\altaffiltext{6}{Dept.~of Physics and Wisconsin IceCube Particle Astrophysics Center, University of Wisconsin, Madison, WI 53706, USA}
\altaffiltext{7}{Oskar Klein Centre and Dept.~of Physics, Stockholm University, SE-10691 Stockholm, Sweden}
\altaffiltext{8}{Erlangen Centre for Astroparticle Physics, Friedrich-Alexander-Universit\"at Erlangen-N\"urnberg, D-91058 Erlangen, Germany}
\altaffiltext{9}{Department of Physics, Marquette University, Milwaukee, WI, 53201, USA}
\altaffiltext{10}{Dept.~of Physics, Pennsylvania State University, University Park, PA 16802, USA}
% ...

\email{analysis@icecube.wisc.edu}

\begin{abstract}

The IceCube Collaboration has previously discovered a high-energy astrophysical neutrino flux using neutrino events with interaction vertices contained within the instrumented volume of the IceCube detector.
We present a complementary measurement using charged current muon neutrino events where the interaction vertex can be outside this volume.
As a consequence of the large muon range the effective area is significantly larger but the field of view is restricted to the Northern Hemisphere.
IceCube data from 2009 through 2015 have been analyzed using a likelihood approach based on the reconstructed muon energy and zenith angle.
At the highest neutrino energies between $194\,\mathrm{TeV}$ and $7.8\,\mathrm{PeV}$ a significant astrophysical contribution is observed, excluding a purely atmospheric origin of these events at $5.6\,\sigma$ significance.
The data are well described by an isotropic, unbroken power law flux with a normalization at $100\,\mathrm{TeV}$ neutrino energy of 
$\left(0.90^{+0.30}_{-0.27}\right) \times 10^{-18}\,\mathrm{GeV^{-1}\,cm^{-2}\,s^{-1}\,sr^{-1}}$ and a hard spectral index of $\gamma = 2.13 \pm 0.13$.
The observed spectrum is harder in comparison to previous IceCube analyses with lower energy thresholds which may indicate a break in the astrophysical neutrino spectrum of unknown origin.
The highest energy event observed has a reconstructed muon energy of $(4.5\pm1.2)\,\unit{PeV}$ which implies a probability of less than $0.005\%$ for this event to be of atmospheric origin.
Analyzing the arrival directions of all events with reconstructed muon energies above $200\,\mathrm{TeV}$ no correlation with known $\gamma$-ray sources was found.
Using the high statistics of atmospheric neutrinos we report the currently best constraints on a prompt atmospheric muon neutrino flux originating from charmed meson decays which is below $1.06$ in units of the flux normalization of the model in \cite{Prompt:ERS}.

\end{abstract}


\section{Introduction}\label{sec:introduction}

The detection of high-energy cosmic neutrinos as 
cosmic messengers has been an important goal of astroparticle physics.
Being stable, electrically neutral particles, high-energy neutrinos are able to propagate almost undisturbed through the universe 
from their production sites to Earth keeping their directional and energy information.
 Hence, they constitute 
 excellent cosmic messenger particles, particularly at the highest energies. They 
 arise from weak 
decays of hadrons, mostly pions and kaons, which are expected to be 
produced by hadronic interactions of cosmic-rays in the surrounding matter of the cosmic-ray accelerator. 
Their observation will help to elucidate the unknown 
sources of high-energy cosmic-rays \citep{Gaisser:Halzen:Stanev,Learned:Mannheim,Becker:2007sv}.

Already in the 1960s the observation of high-energy neutrinos was discussed by \cite{Greisen:1960, Markov:1960vja,Reines:1960}, shortly after the discovery of the neutrino by \cite{Reines:1956}. 
The proposed method was the detection of up-going muons as a signature of 
a charged-current (CC) muon neutrino interaction below the detector. 
Soon it was realized that the expected astrophysical fluxes are small and cubic-kilometer sized detectors would be needed to accomplish the goal, see e.g. \cite{Roberts:1992re}. The construction of large
Cherenkov detectors by instrumenting optically transparent natural media, i.e. 
deep oceans, lakes and glaciers with photo-sensors \citep{Belolaptikov:1997ry,Andres:1999hm,antares:2011nsa} proved to be a key concept.
The largest instrument to date is the IceCube Neutrino Observatory at the geographic South Pole, \cite{IceCube:FirstYear}.

Main backgrounds to the search for astrophysical neutrinos are high-energy
atmospheric neutrinos and muons produced by cosmic-ray 
interactions in the Earth's atmosphere. 

In 2013, a diffuse all-flavor flux of high-energy astrophysical neutrinos was discovered \citep{Aartsen:2013jdh,IceCube:HESE3Years}.
The analysis selected events due to high-energy neutrinos which interact within the detector by using its outer layers as a veto. This strategy enables a full-sky sensitivity for all neutrino flavors. The veto not only rejects atmospheric muons entering the detector from the outside extremely efficiently, but also atmospheric neutrinos from above the detector which are produced together with muons. 

\begin{figure*}
	\centering
	\includegraphics[width=1.\textwidth]{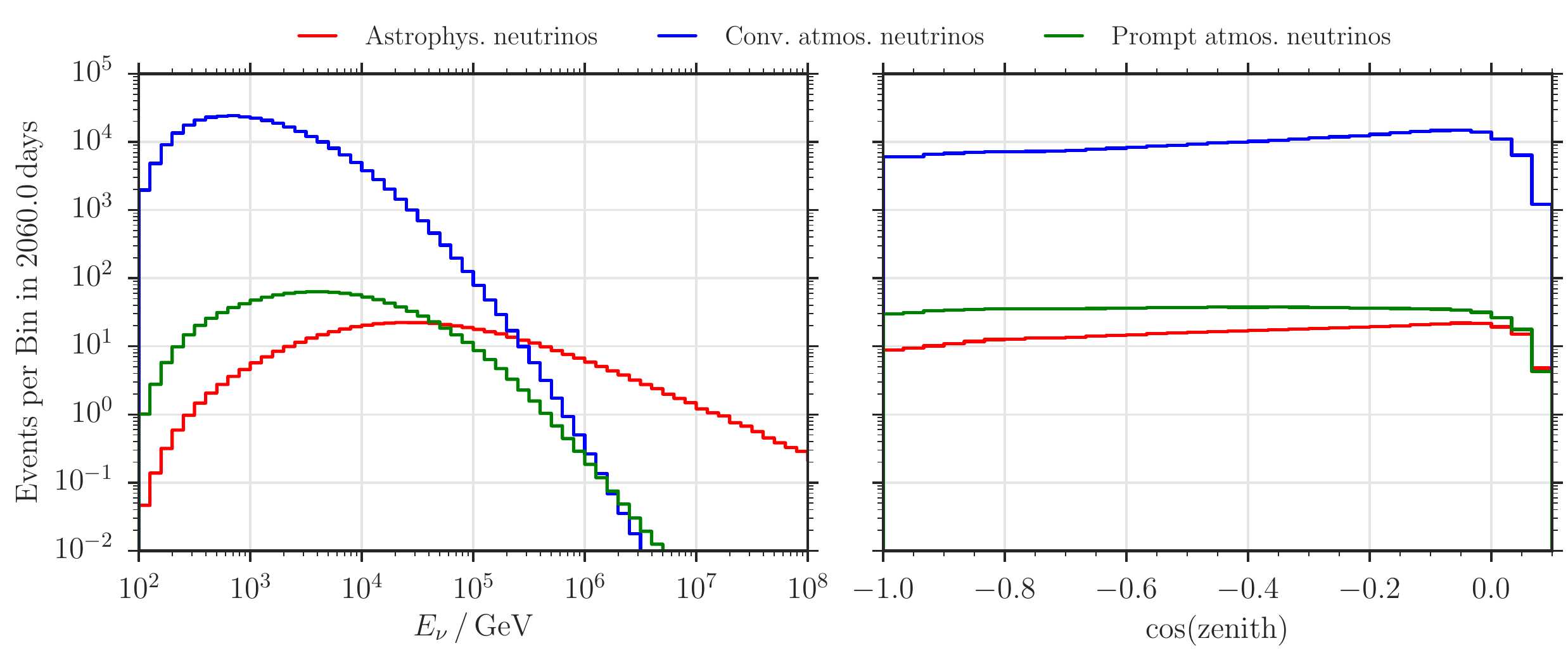}
	\caption{Distribution of the expected neutrino energy (left) and zenith angle (right) for the data selection of this analysis. Shown are the distributions of conventional atmospheric neutrinos \citep{Conv:Honda2007}, prompt atmospheric neutrinos \citep{Prompt:ERS} where both are corrected for the cosmic-ray spectrum in \cite{Gaisser:2012zz} and a benchmark astrophysical signal $10^{-18}\,\mathrm{GeV^{-1}\,cm^{-2}\,sr^{-1}\,s^{-1}} (E_\nu/100\,\mathrm{TeV})^{-2}$. \label{fig:astro_signature}}
\end{figure*}

In this analysis we focus on up-going muons
which arise from charged-current interactions of muon neutrinos
both inside and outside the detector.
By allowing neutrinos to interact outside the instrumented volume a larger effective area is achieved. However, at the same time it is necessary to restrict the analysis to the Northern hemisphere where the Earth filters atmospheric muons efficiently. Furthermore, the analysis is mainly sensitive to a muon neutrino flux because of the large muon range. 
Nevertheless this strategy will impose further constraints on possible models \citep{PhysRevD.87.063011, PhysRevD.88.047301, JCAP.2013.01.028, PhysRevLett.111.041103, JCAP.2012.06.030, PhysRevD.88.043009, PhysRevD.88.081302, GonzalezGarcia201439, PhysRevD.89.083003,  Tamborra:2014:Starforming, MuraseBlazers:2014, Bechtol:2015, Senno:2015} that have been proposed to explain the observed astrophysical neutrino flux. 

This analysis is based on a high-purity and high-statistics selection
of about $350 000$ well-reconstructed up-going muon events from six years of IceCube operation,
improving the statistics compared to previous analyses \citep{IceCube:IC59NuMuDiffuse,IceCube:IC79NuMuDiffuse} by almost an order of magnitude.

Even when individual astrophysical neutrino sources cannot be identified because they are too weak, their cumulative flux can be measured as a diffuse flux. 
The signature of an astrophysical neutrino signal with respect to the background of atmospheric neutrinos
is illustrated in Fig. \ref{fig:astro_signature}.
Astrophysical neutrinos from cosmic accelerators are generically expected to have a hard energy spectrum as originally predicted by Fermi:
$dN_{\nu}/dE \simeq \phi_0 \cdot E^{-2}$. However, the spectral index depends in detail on the source properties and the acceleration mechanism \citep{Bell201356, Kashti:2005qa, Klein:2012ug}. Recent IceCube analyses \citep{Aartsen:2015zva:hese, IceCube:MESE2Years, Aartsen:2015knd, Aartsen:2015zva:casc} yielded a softer spectrum with a spectral index between 2.5 and 2.7.

The energy spectrum of the atmospheric neutrino background is about one power steeper than the primary cosmic-ray spectrum ($dN_{CR}/dE \propto E^{-2.7..3.1} $), 
with the exception of prompt neutrinos from heavy meson decays, 
which follow the 
primary spectrum more closely. The astrophysical signal appears as an excess above
energies of about $100\,\mathrm{TeV}$.
As shown the zenith distribution differs for signal and backgrounds which themselves depend on the energy.
At the highest energies the Earth becomes increasingly opaque to neutrinos and the signal is dominated by events near the horizon.

The identification of an astrophysical signal is based on a two-dimensional likelihood fit in zenith and energy. It follows the methods of the previous analyses \citep{IceCube:IC59NuMuDiffuse,IceCube:IC79NuMuDiffuse} which are improved with respect to the treatment of systematic uncertainties.

The data selection is described in Sec. \ref{sec:data_sample}. The method is described in Sec. \ref{sec:analysis_method}. The results of the analysis with respect to the astrophysical signal are presented in Sec. \ref{sec:results}, where we discuss the fit results, tests of alternative hypotheses and investigations on the most energetic event \citep{schoenen2015detection}. In Sec. \ref{sec:aniso} we present investigations on the directions of recorded events and the attempt to correlate these directions with astrophysical objects. In Sec. \ref{sec:prompt} we discuss implications of this analysis for the expected flux of high-energy prompt atmospheric neutrinos from the decay of charmed mesons and obtain the
currently most constraining exclusion limit.

\section{Data Sample}\label{sec:data_sample}

\subsection{IceCube Detector}
The IceCube Neutrino Observatory is a cubic-kilometer-sized Cherenkov detector embedded in the ice
at the geographic South Pole \citep{IceCube:FirstYear}.
It has been designed to detect neutrinos above TeV energies by measuring the Cherenkov light produced by charged particles produced in neutrino interactions. 
A total of 5160 optical photomultiplier tubes (PMTs) instrument 
 86 cable strings with a vertical spacing of $17\,\mathrm{m}$
at depths between $1450$--$2450$\,m
beneath the surface of the glacial ice sheet \citep{Abbasi:2010vc}. 
 Each PMT is housed in a
digital optical module (DOM), consisting of a pressure-resistant sphere, 
digitization/calibration electronics and calibration LEDs \citep{Abbasi:2008aa}.

The strings are deployed in a hexagonal pattern with an inter-string spacing of about $125\,\mathrm{m}$ except for the central eight strings which 
have a smaller spacing of about $60\,\mathrm{m}$ and also a smaller vertical
 DOM spacing.

The detector was completed in December of 2010; prior to that, data were recorded with partially installed detector configurations.
In the remainder of the paper we differentiate the partial detector configurations by the number of strings, e.g. IC59 for the 59-string configuration. The complete detector with 86 strings is referred to by the year the data taking started, e.g. IC2011. The analysis presented here uses data taken from May 2009 until May 2015 which includes the partial detector configurations IC59, IC79 and the seasons IC2011--2014 of the completed detector.

\subsection{Event Selection}

The events that trigger IceCube dominantly are down-going atmospheric muons produced in cosmic-ray air showers.
The standard trigger condition for high-energy neutrino analyses in IceCube requires a minimum of eight DOMs recording light within a time window of $5\,\mu $s, which results in a rate above $2\,\mathrm{kHz}$.
The triggering DOMs must be in a local coincidence with either their neighboring or next-to-nearest neighboring DOMs.

For each trigger the digitized PMT waveforms of the detected Cherenkov-light signals are sent to the surface where the number of photons as well as the arrival times of photons are extracted. This information is used to reconstruct the energy and geometry of the event \citep{Ahrens:2003fg,EnergyReco:truncated,Aartsen:2013vja}.

The data processing schemes were improved during the construction of 
IceCube and the event selection has been optimized for each detector 
configuration.
Data are processed and reconstructed at the South Pole in real time. A filter criterion, optimized for high-energy track-like signatures, requires a minimum amount of detected total charge and a good quality of the track reconstruction.
This reduces the data stream to about $34\,\mathrm{Hz}$ that is sent off-site via satellite for further data processing.
These events are still dominated by down-going atmospheric muon events. In order to select high-energy up-going muons with high purity and high efficiency, more sophisticated reconstruction algorithms are applied and high quality events are selected.

The neutrino event selections are based on Monte Carlo (MC) generated neutrinos and atmospheric muons. Note that there are differences in the MC used for the different seasons due to improving simulation code and models which is accounted for in the likelihood fit (cf. \ref{sec:systematics}). The simulation of neutrinos is performed by injecting a neutrino at the Earth's surface and propagating it through the Earth.
The neutrino interaction in ice or rock is simulated \citep{ANIS} with the deep inelastic scattering cross section calculated using the CTEQ5 parton distribution functions \citep{CTEQ5} or the updated HERA1.5 PDFs \citep{CSMS}. At the energies of interest the cross sections differ by less than 5\%.
Each simulated neutrino is forced to interact in the vicinity of the instrumented volume. The volume is scaled as a function of the neutrino energy to include the maximum range of the muon produced in the interaction. The muon is propagated through the detector taking into account energy losses and decay \citep{MMC, PROPOSAL}. The Cherenkov light from charged particles is tracked through the ice to the DOMs \citep{Lundberg:PHOTONICS, Chirkin:PPC, Kopper:CLSIM} taking into account the Antartic ice properties \citep{ackermann2006optical, IceCube:OpticalIcePropertiesPapers,Aartsen:2013ola:chirkin}. Lastly, the detector response and data acquisition are simulated. The same simulation chain is used for atmospheric muons which are simulated with CORSIKA \citep{Heck:1998vt}. Both, neutrino simulation and atmospheric muon simulation, can be weighted to different fluxes.

The event selection for IC59 is identical to \cite{IceCube:IC59NuMuDiffuse} and covers the up-going zenith range $90^\circ$ -- $180^\circ$.
For the later seasons the zenith range has been enlarged to additionally cover angles between $85^\circ$ and $90^\circ$ as in \cite{IceCube:IC79NuMuDiffuse} where the overburden by the antarctic ice sheet is still more than $12\,\mathrm{km}$ of water equivalent.
Additionally, the separation of mis-reconstructed atmospheric muons and well reconstructed neutrino induced muons has been improved by using boosted decision trees (BDT). For IC2011 and later the AdaBoost algorithm \citep{Freund:1997xna} implemented in \cite{pedregosa2011scikit} has been used.
Due to filter and processing changes after the first complete detector season a separate BDT has been trained for IC2011 and the data are treated separately as for the seasons IC59 and IC79.

For the optimization of the BDT we use simulations of up-going muon neutrinos following an $E^{-2}$ spectrum which produce a muon via a charge-current interaction. 
In addition, to define the signal for the BDT, only simulated events with directions reconstructed to better than $5^\circ$ are used.
The background is defined by atmospheric muons from cosmic-ray air-showers
that have been mis-reconstructed as up-going. 
The simulation is weighted to the cosmic-ray model in \cite{Gaisser:2012zz}.
The features used in the training of the BDT are characteristics of the event topology and parameters evaluating the quality of the reconstructions.
These parameters have been selected requiring good agreement between experimental and simulated data.
The threshold of the BDT classifier is chosen by considering the neutrino selection efficiency and the purity. 
In order to model the atmospheric background using simulation a high purity is required rejecting nearly all atmospheric muons.
The chosen threshold results in a purity which is better than $99.7\%$. The remaining background clusters at low-energies and is strongly dominated by atmospheric neutrinos. 
Thus, it cannot affect the analysis and therefore does not have to be taken into account as a separate template in the likelihood fit.
The performance estimates are based on 10-fold cross validation \citep{narsky2013statistical} and a separate validation set.
Additionally, a fit of the data has been performed excluding events from above the horizon between $85^\circ$ and $90^\circ$.
Since the fit results remain nearly unaffected we conclude that the fit isn't biased by any unaccounted high-energy muons.

\begin{figure}[!h]
	\centering
	\includegraphics[width=0.98\columnwidth]{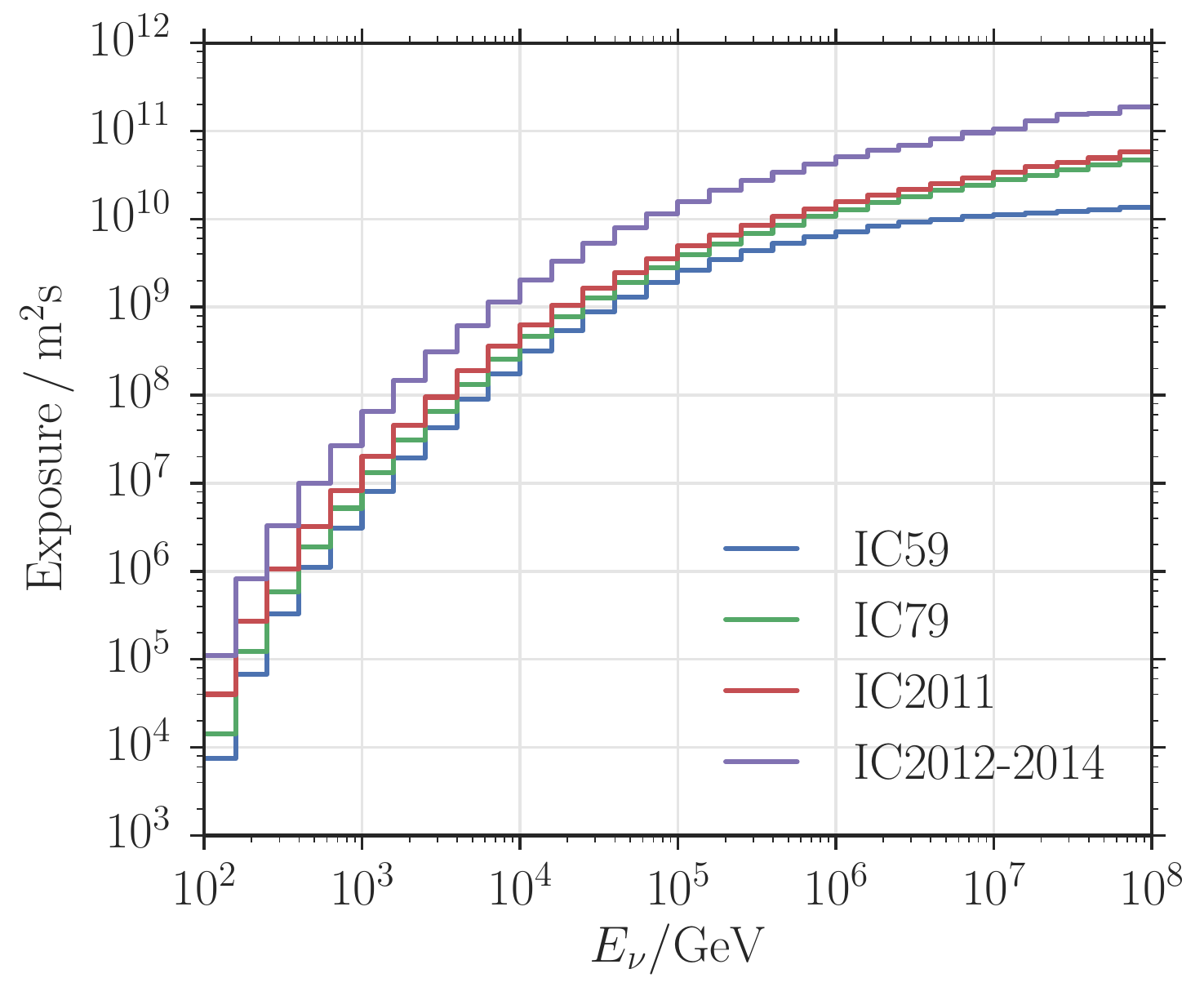}
	\includegraphics[width=0.98\columnwidth]{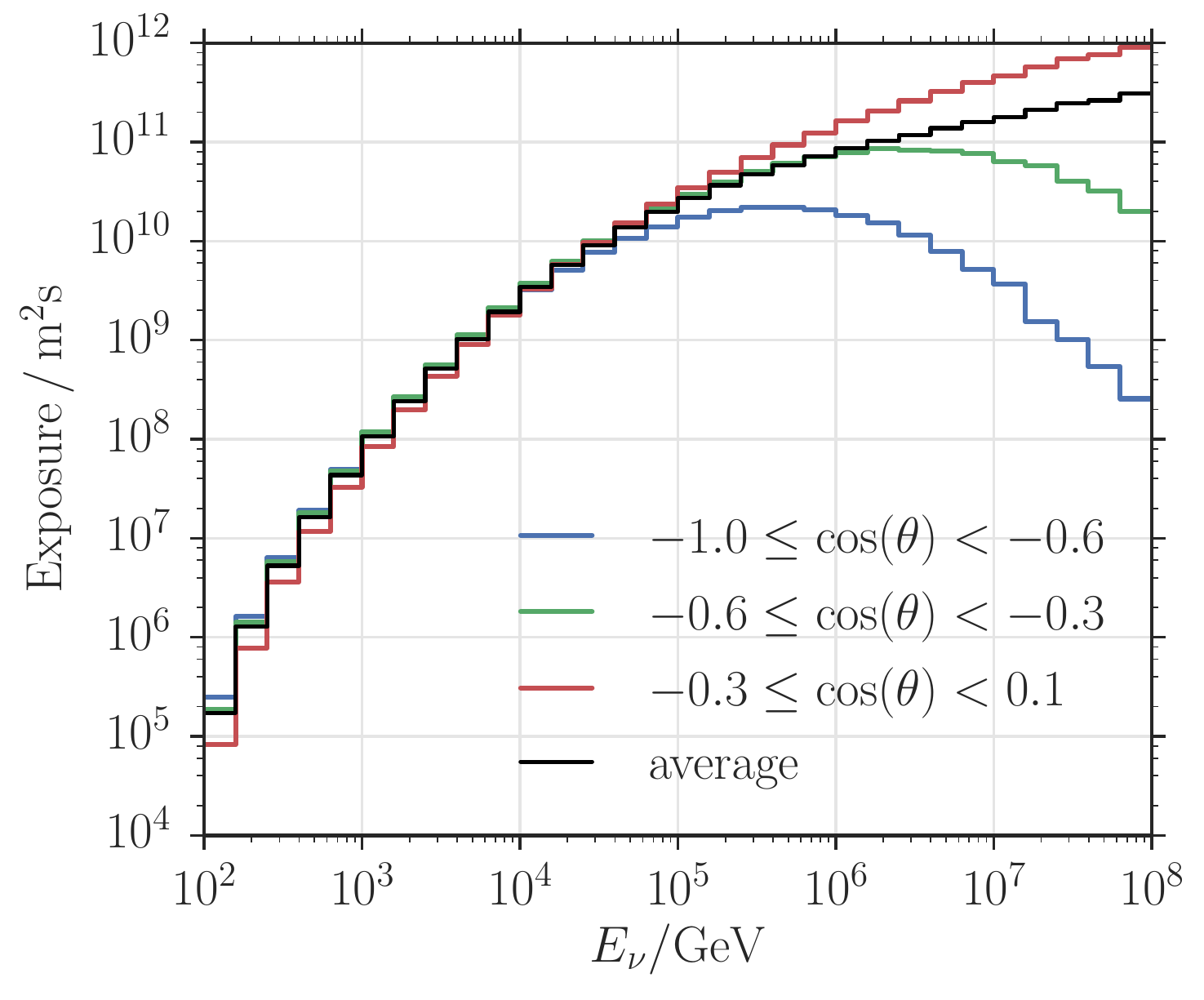}
	\caption{Exposure. Top: Individual contributions to the total exposure from the different detector configurations. Bottom: Total exposure for different zenith regions for the combined dataset. Note that the exposure is based on the sum of the effective areas of $\nu_\mu$ and $\bar{\nu}_\mu$. Therefore, the total number of events is obtained by integrating the product of exposure and averaged neutrino flux $\phi_{\nu_\mu+\bar{\nu}_\mu} / 2$ over the neutrino energy and solid angle. \label{fig:exposure}}
\end{figure}

Figure \ref{fig:exposure} shows the total exposure for the different detector configurations and for the full dataset for different ranges in cosine zenith.
The total number of events as well as the total live time categorized by season are summarized in Tab. \ref{tab:event_selection}.
For the best-fit astrophysical flux (cf. Sec. \ref{sec:results:astroflux}) the expected number of astrophysical muon neutrinos included in these data is approximately $500$.

\begin{deluxetable}{lccc}
  \tablecaption{Summary of the data selection. 
  The table gives the number of events and the effective live time for each used dataset. \label{tab:event_selection}}
  \tablehead{\colhead{Season} & \colhead{$\theta_\mathrm{min}-\theta_\mathrm{max}$ (deg)} & \colhead{$t_\mathrm{live}$ (days)} & \colhead{$N_\mathrm{event}$}}
  \startdata
    IC59 & 90--180 & 348.1  & 21,411 \\
    IC79 & 85--180 & 310.0 & 36,880 \\
    IC2011 & 85--180 & 342.1 & 71,191 \\
    IC2012--2014 & 85--180 & 1059.8 & 222,812 \\
  \enddata
\end{deluxetable}

\section{Analysis Method}\label{sec:analysis_method}

\subsection{Likelihood method}
The experimental and simulated data are binned in two observables sensitive to distinguish between signal and background: reconstructed muon energy and cosine of the zenith angle. These bins are 
analyzed by a maximum likelihood approach. 
The expectation in each bin is a function of the signal and nuisance parameters.
The likelihood used in this analysis is given by \cite{LLH:Dima} and 
is the same as used in \cite{IceCube:IC59NuMuDiffuse}. The likelihood per bin is
\begin{equation}
  \mathcal{L}_i = \left( \frac{\mu_i}{s_i/n_\mathrm{s}}\right)^{s_i} \cdot \left( \frac{\mu_i}{d_i}\right)^{d_i}, \label{Eq:likelihood}
\end{equation}
where $n_\mathrm{s}$ defines the ratio of the live times for simulation and experimental data, 
$d_i$ is the number of events in data, $s_i$ the number of simulated events and $\mu_i$ the expectation in bin $i$.
With this likelihood the expectation $\mu_i$ is optimized based on the knowledge about the statistics of the simulated and experimental dataset.
Unlike a Poisson likelihood, we account for the finite 
statistics of simulated data which becomes relevant for small bin contents
in a multi-dimensional parameter space.
In the limit of infinite statistics of simulated data this likelihood converges to a saturated Poisson likelihood.
A version of Eq. \ref{Eq:likelihood}, modified for weighted events according to \cite{LLH:Dima}, is used in the analysis.

The per-bin expectation is given by
\begin{eqnarray*}
  \mu_i\left(\boldsymbol{\theta};\boldsymbol{\xi}\right) & = &\mu_i^\mathrm{conv.}\left(\boldsymbol{\xi}\right)\\
							 & + &\mu_i^\mathrm{prompt}\left(\Phi_\mathrm{prompt};\Delta\gamma_\mathrm{CR},\lambda_\mathrm{CR},\boldsymbol{\xi}_\mathrm{det}\right)\\
							 & + &\mu_i^\mathrm{astro.}\left(\Phi_\mathrm{astro},\gamma_\mathrm{astro};\boldsymbol{\xi}_\mathrm{det}\right),
\end{eqnarray*}
which depends on the signal $\boldsymbol{\theta}$ and nuisance parameters $\boldsymbol{\xi}$ (cf. Tab. \ref{tab:nuisance_parameters}).
Here, $\boldsymbol{\xi}_\mathrm{det}$ corresponds to the parameters taking into account the neutrino detection uncertainties (cf. Sec. \ref{sec:nu_detection_uncert}).
In this analysis, the signal parameters consist of the astrophysical flux parameters and the prompt flux parameter.
The astrophysical flux model used here is a single power law flux described by two parameters: the normalization $\Phi_\mathrm{astro}$ at $100\,\mathrm{TeV}$ neutrino energy and the spectral index $\gamma_\mathrm{astro}$:
\begin{equation}
  \Phi_{\nu+\overline{\nu}} = \Phi_\mathrm{astro} \cdot \left(\frac{E}{100\,\mathrm{TeV}}\right)^\mathrm{-\gamma_\mathrm{astro}}.
\end{equation}
The prompt neutrino flux is described by the prediction taken from \cite{Prompt:ERS} (ERS) where the absolute normalization $\Phi_\mathrm{prompt}$ is taken as free parameter.
In addition, nuisance parameters are introduced to take into account systematic uncertainties,
e.g. the conventional atmospheric neutrino flux is described by the prediction taken from \cite{Conv:Honda2007} and the flux normalization $\Phi_\mathrm{conv}$ is taken as a nuisance parameter.
Note that the conventional and prompt neutrino flux predictions have been corrected for the knee in the cosmic-ray spectrum based on the cosmic-ray models in \cite{hoerandel2003knee} and \cite{Gaisser:2012zz} (cf. Sec. \ref{sec:systematics}).
The implementation of nuisance parameters is described in Sec. \ref{sec:systematics} in more detail.

The global likelihood, which is maximized, is the product of all per-bin-likelihoods $\mathcal{L}=\prod_i\mathcal{L}_i$.
The significances and parameter uncertainties in this analysis are derived using the profile likelihood technique and Wilks' theorem \citep{WilksTheorem}.
The applicability of Wilks' theorem has been tested and verified by ensemble studies.

\subsection{Systematic uncertainties} \label{sec:systematics}
In order to account for systematic uncertainties, resulting from the imperfect background and signal modeling, continuous nuisance parameters valid for the entire energy and zenith range are introduced.
The systematic uncertainties can be divided into two categories: neutrino detection uncertainties and atmospheric flux uncertainties. The former include the optical efficiency of the detector, the neutrino-nucleon cross section, the muon energy loss cross section and the optical properties of the Antarctic ice. The latter include the
flux normalizations, the spectral shape and composition of the cosmic-ray spectrum in the ``knee'' region, the spectral index of the primary cosmic-ray spectrum and the relative production yield of
pions and kaons in the atmosphere. The implementation of these uncertainties as nuisance parameters in the likelihood function is done similar to \cite{IceCube:IC59NuMuDiffuse}.
Main improvements with respect to previous analyses
are the parameterizations of the systematic detector effects as unbinned
functions of both fit observables \citep{PhDThesis:Schoenen} using adaptive
kernel density estimation 
and the interpolations between specific models to account for the model uncertainties. 
The systematic detector effects are studied by simulated datasets where the default parameters are changed within their uncertainties.
In cases where no interpolation between specific models is used, the nuisance parameters are implemented
by using independent correction factors $f_k(\xi_k)$. These factors scale the default per-bin expectation $\mu^0_i$ for each flux contribution with respect to the individual nuisance parameter $\xi_k$:
\begin{equation*}
  \mu^0_i \mapsto \mu^0_i \cdot \prod_k f_k(\xi_k).  
\end{equation*}

\subsubsection{Neutrino detection uncertainties}\label{sec:nu_detection_uncert}
\paragraph{Optical efficiency of the detector}
The optical efficiency $\epsilon_\mathrm{opt}$ takes into account all uncertainties related to the light production and detection in the detector, e.g. the number of produced Cherenkov photons, the overall optical transparency of the ice, 
the photon detection efficiency of the digital optical modules and the shadowing of photons by detector components. Since the optical efficiency is directly connected to the brightness of an event as observed with the detector its uncertainty results in an uncertainty on the reconstructed energy scale. The effect has been parametrized as a function of the muon energy proxy and the cosine zenith angle and is implemented as a nuisance parameter. The uncertainty on the optical efficiency is estimated to be less than 15\%. Since the ice properties of the refrozen water within the drill holes differ from the bulk ice properties they are taken into account as a modification of the angular acceptance \cite{IceCube:OpticalIcePropertiesPapers}.

\paragraph{Optical properties of the Antarctic ice}
The probability of a Cherenkov photon to be detected by a digital optical module depends not only on the optical efficiency of the detector but also on the optical transparency of the Antarctic ice.
The main processes are scattering and absorption of photons on their path to the digital optical module. For the Antarctic ice this is modeled by depth-dependent scattering and absorption lengths.
The modeling is done by using measured data from calibration light sources that are integrated into the digital optical modules. Different models
of the ice have been developed during the operation of IceCube. For this analysis the following ice models are used: \textit{WHAM} for IC59 (based on a measurement of the optical properties of the glacial ice at the South Pole presented in \cite{ackermann2006optical}), \textit{SpiceMie} for IC59, IC79 and IC2011 \citep{IceCube:OpticalIcePropertiesPapers} and \textit{SpiceLea} for IC2011-2014 \citep{Aartsen:2013ola:chirkin}.
For all available simulation datasets, the effects of the optical ice properties as a function of reconstructed energy proxy and cosine zenith are parametrized and implemented as a nuisance parameter. This is done for different ice models and each detector configuration.
The parameterization is done by introducing a parameter $\lambda_\mathrm{ice}$ that describes a linear combination ($\lambda_\mathrm{ice}\cdot\mathrm{M_1} + (1-\lambda_\mathrm{ice})\cdot\mathrm{M_2}$) between two ice models where $\mathrm{M_1}$ and $\mathrm{M_2}$ are the expectations per bin corresponding to the two ice models. In addition, for a given ice model the effect of different scattering lengths $\lambda_\mathrm{scat}$ and absorption lengths $\lambda_\mathrm{abs}$ on the muon energy proxy and the cosine zenith angle have been parameterized. Because of missing simulations this could not be done for IC59. For IC79 the scattering and absorption lengths have been varied simultaneously resulting in only one effective nuisance parameter $\lambda_\mathrm{abs/scat}$. From IC2011 on the scattering and absorption lengths have been varied separately. For more information see Tab. \ref{tab:nuisance_parameters}. The individual uncertainty for both quantities is estimated to be less than $10$\,\%. 
The scattering length mainly influences the angular resolution of the neutrino arrival direction and therefore the reconstructed zenith angle. Since the cosine zenith bin width of the analysis is relatively coarse, the effect of this uncertainty on the observable distribution is small. The absorption length mainly influences the flux normalization and the shape of the energy distribution. This effect is much larger, compared to the scattering length effect.

\subsubsection{Atmospheric flux uncertainties}
\paragraph{Flux normalization}
The uncertainty on the normalization of the conventional atmospheric neutrino flux is implemented as a nuisance parameter $\Phi_\mathrm{conv}$ that scales
the flux normalization of the model by \cite{Conv:Honda2007}. This
model has been extrapolated to higher energy based on the method in
taking into account a more realistic spectrum of cosmic-rays and their composition \citep{Illana:2010gh,AnnePhD}.
Note that the
uncertainty of this parameter is relatively large, on the order of 30\%.
Thus, it absorbs any kind of uncertainty which influences the global flux normalization in the fit.

\paragraph{Cosmic-ray model and spectral index}
The composition of the cosmic-rays is uncertain, in particular above the knee at
 an energy of about $3\,\mathrm{PeV}$. 
Models are based on the superposition of galactic cosmic-rays with rigidity-dependent cut-offs and an emerging extragalactic component. Since conventional and prompt atmospheric neutrinos are produced by cosmic-ray interactions within the atmosphere, the uncertainty on the cosmic-ray spectrum also affects the expectation of these neutrinos. The effect of different cosmic-ray models is parameterized by $\lambda_\mathrm{CR}$ as a function of the muon energy proxy and the cosine zenith angle similar to the discrete ice models. Here, a linear combination between \cite{hoerandel2003knee} and \cite{Gaisser:2012zz}, which are the extreme cases, is used.
In addition to the effects between different cosmic-ray models an overall change in the cosmic-ray spectral index affects the expectation of atmospheric neutrinos. Therefore, a shift of the cosmic-ray spectral index $\Delta\gamma_\mathrm{CR}$ is implemented as a nuisance parameter representing the uncertainty on the cosmic-ray spectral index. The uncertainty is estimated to be of the order of 4\% based on differences between the aforementioned cosmic-ray models. A positive $\Delta\gamma_\mathrm{CR}$ corresponds to a softer energy spectrum.

\paragraph{Kaon-to-pion ratio}
Conventional atmospheric neutrinos are produced by decays of pions and kaons which are themselves produced in air showers.
The relative contribution of kaons and pions $\mathrm{K}/\pi$ to the production of conventional atmospheric neutrinos affects their zenith angle distribution. 
In this analysis it is defined by the ratio of the integrated neutrino fluxes from kaon and pion decays. Using the neutrino flux parameterization from \cite{GaisserTextbook} fitted to the conventional atmospheric neutrino prediction from \cite{Conv:Honda2007} between $1-10\,\mathrm{TeV}$, the kaon-to-pion ratio is implemented as a nuisance parameter where the uncertainty is estimated to be of the order of 10\%.

\paragraph{Atmospheric temperature effects}
The expected number of conventional atmospheric neutrinos is directly connected to the number of pion and kaon decays in the atmosphere. A denser atmosphere will increase the interaction probability for pions and kaons relative to the decay probability, which reduces the overall neutrino flux. Thus, annual temperature fluctuations influence the expectation of conventional atmospheric neutrinos. Since the prediction of conventional atmospheric neutrinos from \cite{Conv:Honda2007} is based on the standard US atmosphere, the expectation is corrected for annual temperature fluctuations. This is done using the formalism reported in \cite{ICRC:TemperatureVariation} and data measured by the instrument AIRS installed on the AQUA satellite \citep{AQUA}. The effect of this correction is estimated to be of the order of 2\% with an uncertainty of about 0.1\%.\\

\begin{deluxetable*}{lcccccc}
  \tablecaption{Nuisance parameters. Columns two to five show the best fit values for each data sample individually where the fit was performed on data within a predefined background region. The background regions were defined as follows: for IC59 the muon energy loss proxy must be less than $1\,\mathrm{GeV/m}$ and for IC79, IC2011 and IC2012-2014 the muon energy proxy must be less than $10\,\mathrm{TeV}$. These best fit values are used as default values to define a common baseline. In the combined likelihood fit the default values are then scaled by global nuisance parameters where the best-fit values including the $68\%\,\mathrm{C.L.}$ error determined by the profile likelihood technique are shown in the last column. Column six shows if the scaling is an absolute or relative change with respect to the default values. Note that the nuisance parameters quoted here are allowed to change for each data set to absorb differences in the simulations which are caused by improvements in the simulation code and models. Thus, they do not have to reflect their real physical quantities. \label{tab:nuisance_parameters}}

  \tablehead{\colhead{} & \colhead{IC59} & \colhead{IC79} & \colhead{IC2011} & \colhead{IC2012-2014} & \colhead{scaling} & \colhead{Best Fit ($68\%\,\mathrm{C.L.}$)}}

  \startdata
  \underline{Flux properties:}        &     &       &     &       &     & \\
  Conventional flux $\Phi_\mathrm{conv}$      & 1.028   & 1.047     & 1.184   & 1.194     & relative  & $0.998 \pm 0.003$ \\
  Kaon-pion ratio $\mathrm{K}/\pi$        & 1.310   & 1.514     & 1.002   & 1.032     & relative  & $0.977 \pm 0.027$ \\
  Cosmic-ray spectral index $\Delta\gamma_\mathrm{CR}$  & -0.049  & -0.049    & -0.061  & 0.012     & absolute  & $0.008 ^{+0.004}_{-0.023}$ \\
  Cosmic-ray model $\lambda_\mathrm{CR}$      & 1.0 (H3p)   & 1.0 (H3p)   & 1.0 (H3p)   & 1.0 (H3p)     & absolute  & $0.0 - 0.5$ \vspace{0.2cm}\\
  Optical efficiency $\epsilon_\mathrm{opt}$    & 1.011   & 0.974     & 1.042   & 1.056     & relative  & $1.002 \pm 0.002$ \vspace{0.2cm}\\
  \underline{Optical ice properties:}     &     &       &     &       &     & \\
  Scattering length $\lambda_\mathrm{scat}$     & \nodata    & \nodata      & 1.027   & 1.014     & relative  & $0.999 \pm 0.005$ \\
  Absorption length $\lambda_\mathrm{abs}$      & \nodata    & \nodata      & 1.000   & 1.047     & relative  & $1.001 \pm 0.004$ \\
  Absorption/scattering length $\lambda_\mathrm{abs/scat}$  & \nodata    & 0.991     & \nodata    & \nodata      & relative  & $1.000 \pm 0.004$ \\
  Ice model $\lambda_\mathrm{ice1}$ [SpiceMie,WHAM]   & 0. (SpiceMie) & SpiceMie (fixed)  & \nodata    & \nodata      & absolute  & $0.000 + 0.014$ \\
  Ice model $\lambda_\mathrm{ice2}$ [SpiceMie,SpiceLea] & \nodata    & \nodata      & 0.551   & SpiceLea (fixed)  & absolute  & $0.006 \pm 0.057$ \\
  \enddata
\end{deluxetable*}

\begin{figure}[!h]
	\centering
	\includegraphics[width=\columnwidth]{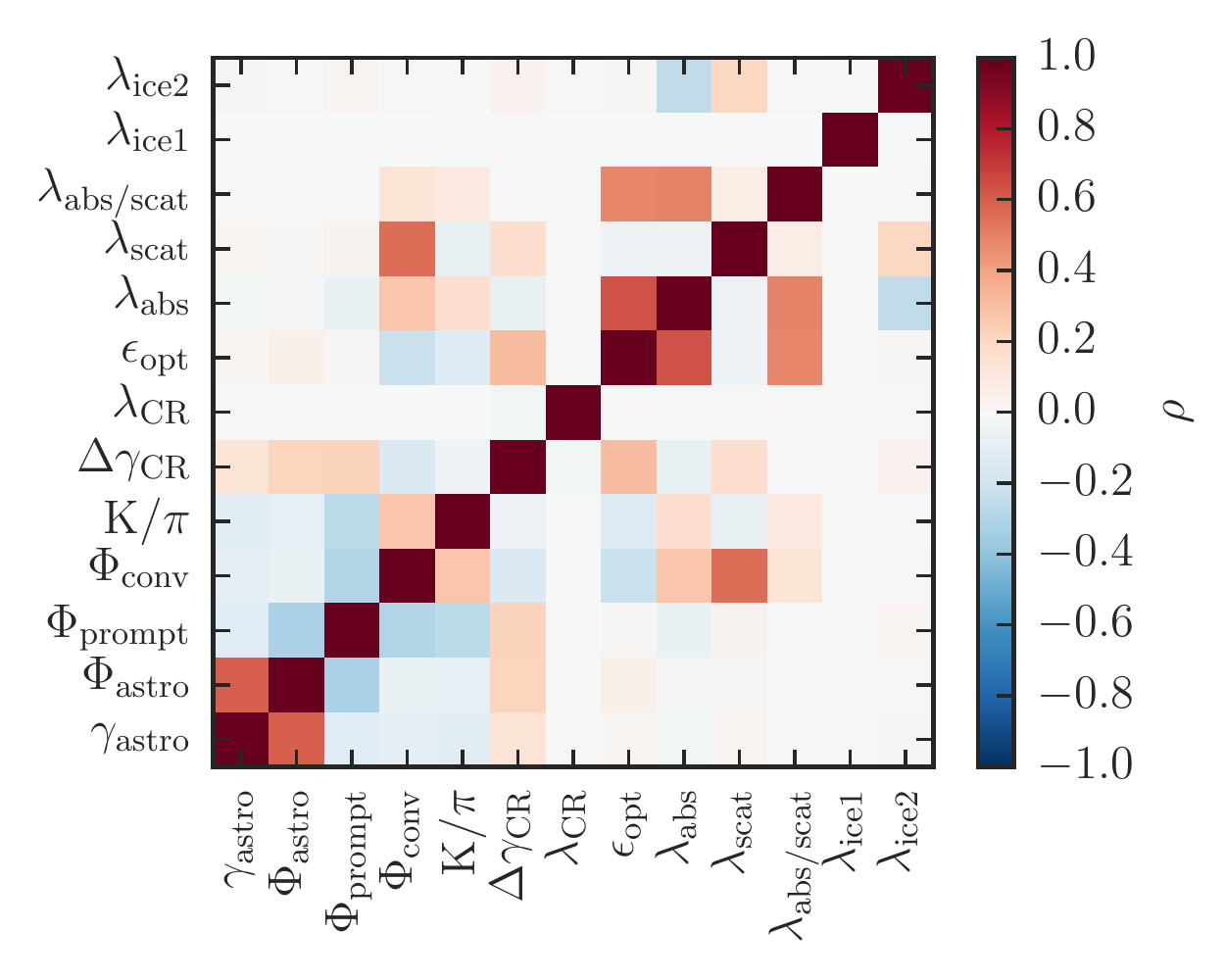}
	\caption{Correlation matrix of signal and nuisance parameters. $\lambda_\mathrm{ice1}$ describes the linear combination between SpiceMie \citep{IceCube:OpticalIcePropertiesPapers} and WHAM \citep{ackermann2006optical} for IC59, $\lambda_\mathrm{ice2}$ describes the linear combination between SpiceMie \citep{IceCube:OpticalIcePropertiesPapers} and SpiceLea \citep{Aartsen:2013ola:chirkin} for IC2011. $\lambda_\mathrm{abs/scat}$ describes the relative change of the ice properties for IC79 and $\lambda_\mathrm{abs}$, $\lambda_\mathrm{scat}$ describes the relative change of the ice properties for IC2011-14. \label{fig:correlation_matrix}}
\end{figure}

With the goal of achieving a unbiased result for the signal parameters, we note that many nuisance parameters were deliberately chosen correlated 
(see Fig. \ref{fig:correlation_matrix}).
For example, the optical detector efficiency is correlated to the Cherenkov light yield uncertainty and to the effects of the uncertainties in the muon energy loss cross sections. 
In cases where the effects of one uncertainty 
are fully absorbed by other nuisance parameters only one parameter has been implemented for better numerical stability of the fit. A list of the implemented parameters is given in Tab. \ref{tab:nuisance_parameters}.
In order to obtain an unbiased result the nuisance parameters are implemented without priors which is tested to have no effect on the sensitivity for an astrophysical or prompt flux.

The fit procedure was tested by generating pseudo experiments where the input parameters were varied. The fit of the signal parameters was found to be robust and unbiased against the choice of nuisance parameters.

\begin{figure*}
	\centering
	\includegraphics[width=1.0\textwidth]{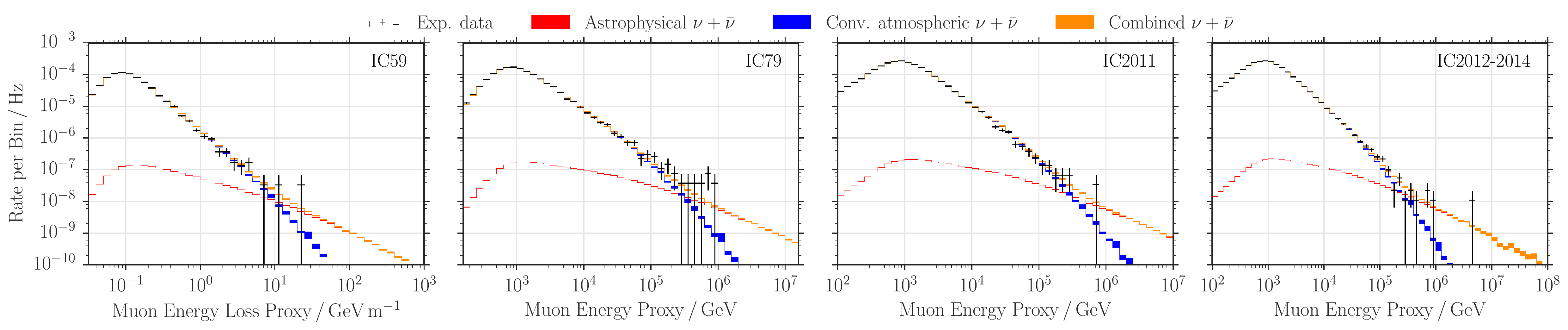}
	\includegraphics[width=1.0\textwidth]{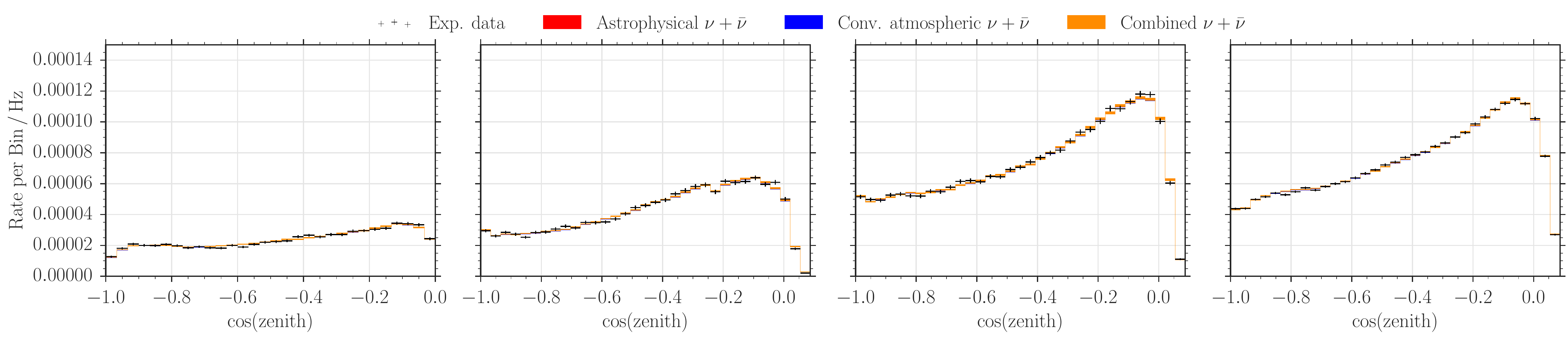}
	\caption{Distributions of the experimental data for
 the muon energy proxy (top) and reconstructed zenith (bottom) for 
each contributing data sample (left: 59-string, center-left: 79-string, center-right: 86-string (2011-2012), right: 86-strings (2012-2015)). Note that IC2011 is different from the later years due to changes in the data processing. The best-fit model for astrophysical and atmospheric neutrinos is superimposed.
Only statistical errors are shown. \label{fig:energy_proj}}
\end{figure*}

Since IceCube's Monte Carlo simulations have evolved and improved 
from year to year, the default expectations
for each nuisance parameter have changed for 
 simulated datasets year by year.
In order to avoid a tension in the fitted nuisance parameters 
induced purely by the differences of the simulations, 
two methods were tested:
the implementation of individual nuisance parameters for each year and the alignment of all nuisance parameters to a common baseline. We found that the two methods give similar sensitivities both for the astrophysical and the prompt flux parameters, and the alignment method is chosen since the time consumption for the fit is much lower. 
The alignment is done by fitting the nuisance parameters in a predefined background region for each year individually.
 The resulting best-fit values for the nuisance parameters, 
summarized in Tab. \ref{tab:nuisance_parameters}, define the default 
values for each year. In the combined likelihood fit of all six 
years these default values are then scaled by global nuisance parameters. 
The scaling can be either an absolute or relative change with respect to the aligned default values.

The flavor composition at Earth is not identical to the flavor composition at an astrophysical source due to neutrino oscillations. For a source, dominated by $p\gamma/pp$-interactions, the initial flavor ratio of 1:2:0 ($\nu_e$:$\nu_\mu$:$\nu_\tau$) is transformed to be approximately equally partitioned among the flavors \citep{Learned:Oscillation,Athar:Oscillation}. Therefore, it is necessary to take into account the muonic decay of taus originating from charged current $\nu_\tau$ interactions by combining it with the astrophysical muon neutrino flux. A fit of the data with and without this contribution shows that accounting for it leads to a decrease in the astrophysical normalization of about $5\%$ and has no effect on the spectral index. In the rest of the paper we account for the contribution of $\nu_\tau$ assuming the equal partitioning.
The astrophysical normalization for other expected flavor compositions of astrophysical neutrino sources can be obtained by rescaling.

\subsection{Parametric Unfolding} \label{sec:analysis_method:unfolding}
The best-fit result for the neutrino energy spectrum measured by this analysis can be used to determine for each event a neutrino energy probability density function $P(E_\nu | E^i_\mathrm{reco})$ with respect to its muon energy proxy $E^i_\mathrm{reco}$. These functions depend on the assumption for the neutrino energy spectrum and are therefore model-dependent. In particular, for the astrophysical neutrino flux an unbroken, single power law is assumed.
For the full six-year data sample the neutrino energy distribution is given by the sum over the probability density function of all events $\sum_i P(E_\nu | E^i_\mathrm{reco})$ where each function is normalized to an event count of one.
Since this approach is model-dependent it is 
called \textit{parametric unfolding} in the following. 
Note that this method cannot replace a model-independent unfolding as done in \cite{Aartsen:2015zva:unfolding} for different IceCube data samples.

\section{Results of the spectral Likelihood fit}\label{sec:results}
\subsection{Fit result}
The result of the fit is presented as a set of one-dimensional projections of 
energy and zenith in Fig. \ref{fig:energy_proj}
separately for each contributing data sample. The experimental data is shown as black crosses, and the best-fit expectations for astrophysical and conventional atmospheric neutrinos are shown as red and blue bands, respectively. 
An excess of high-energy events consistent with an astrophysical signal
above the atmospheric background is visible 
for each data sample. The overall agreement between the data and the MC of the full dataset
 is good for all energies and zenith angles.
We have tested quality of the fit based on the two-dimensional distributions using the ratio between the likelihood (eq. \ref{Eq:likelihood}) and the saturated likelihood as test-statistic \citep{pdg:Agashe:2014kda}. The test-statistic distribution was generated via pseudo experiments based on the best-fit.
The resulting p-value is 95.4\%, indicating a very good agreement between data and MC.
All nuisance parameters are fitted to values consistent with their uncertainty (Tab. \ref{tab:nuisance_parameters}).

Note that the data sample taken with the 79-string configuration contains roughly twice as many high-energy events above a reconstructed muon energy of $100\,\mathrm{TeV}$ than other years (cf. Fig. \ref{fig:energy_proj}).
Nevertheless, the result of the fit for all years is consistent with the fits for individual years.
Visual inspection and other cross checks of these events revealed no indication of any time dependent detector effects. 
A dedicated analysis searching for time dependent neutrino sources \citep{IceCube:TD:PS:2008-2012} has found no indication of a signal. Therefore, the observations are consistent with a statistical fluctuation.
\clearpage

\subsection{Astrophysical flux} \label{sec:results:astroflux}
\begin{figure}
	\centering
	\includegraphics[width=1.\columnwidth]{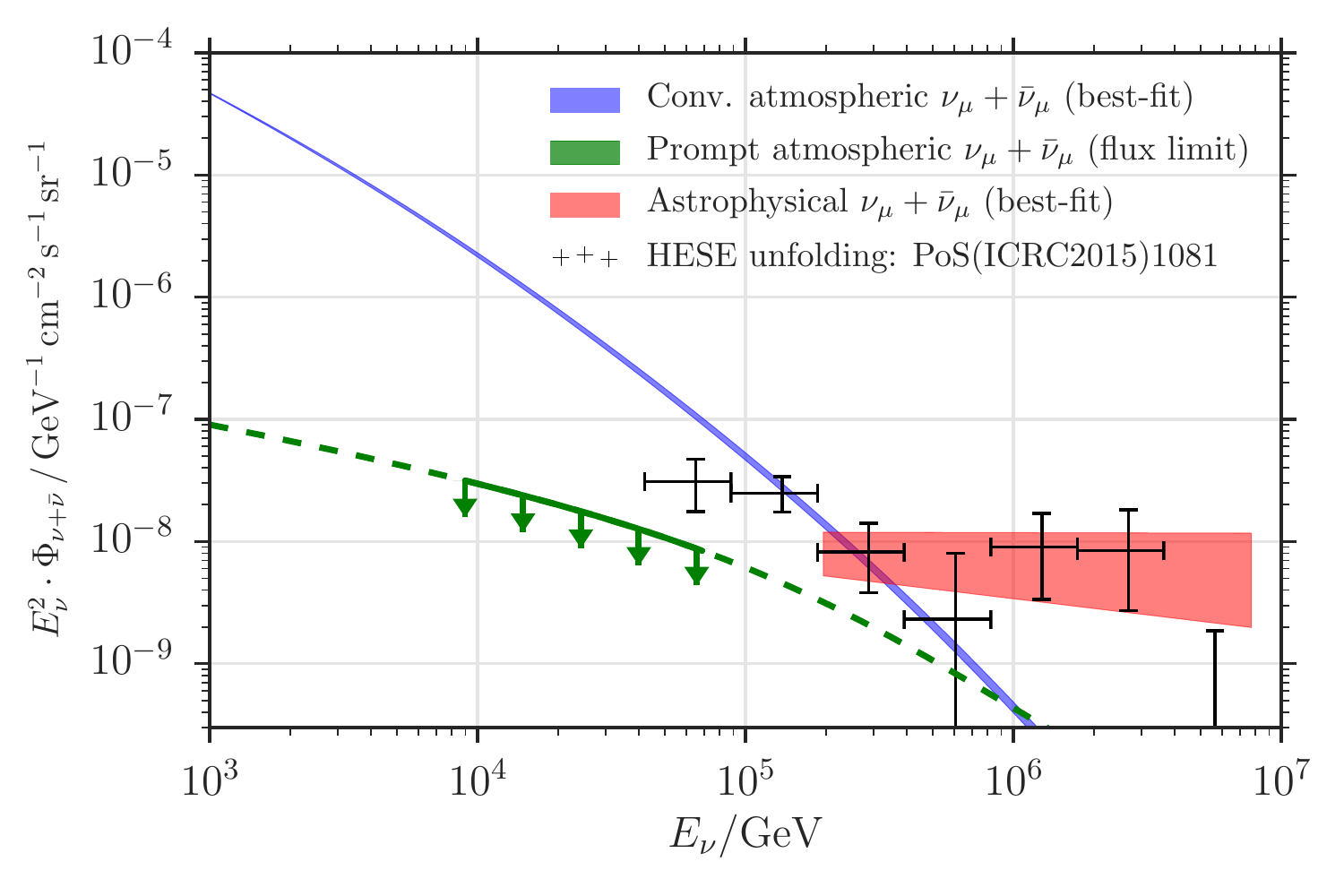}
	\caption{Best-fit neutrino spectra for the unbroken power-law model. 
	The width of the line corresponding to conventional atmospheric neutrinos (blue) represents the one sigma error on the measured spectrum.
	The width of the line corresponding to astrophysical neutrinos (red) shows the effect of varying both astrophysical parameters within one sigma of the best fit values, without accounting for correlation.
	The green line represents the upper limit on the prompt model \citep{Prompt:ERS}.
	The horizontal width of the red band denotes the energy range of neutrino energies which contribute $90\%$ to the total likelihood ratio between the best-fit and the conventional atmospheric-only hypothesis. The black crosses show the unfolded spectrum published in \cite{Aartsen:2015zva:hese}. \label{fig:bestfit_fluxes}}
\end{figure}

The best-fit for the unbroken power-law model of 
the astrophysical flux results in
\begin{equation}
	\Phi_{\nu+\overline{\nu}} = \left(0.90^{+0.30}_{-0.27}\right) \cdot (E_\nu/100\,\mathrm{TeV})^{-(2.13 \pm 0.13)}
\end{equation}
in units of $10^{-18}\,\mathrm{GeV^{-1}\,cm^{-2}\,sr^{-1}\,s^{-1}}$.
The statistical significance of this flux with respect to the atmospheric-only
hypothesis is $5.6$ standard deviations. 
The fit results are shown in Fig. \ref{fig:bestfit_fluxes} and summarized in Tab. \ref{tab:results_physics}.
The quoted errors are based on the profile likelihood using Wilks' theorem \citep{WilksTheorem} and include both statistical and systematic uncertainties.
No contribution from prompt atmospheric neutrinos is preferred by 
the best-fit spectrum and an upper limit, based on the profile likelihood is shown in Fig. \ref{fig:bestfit_fluxes}. For more information about the upper limit for prompt atmospheric neutrinos see Sec. \ref{sec:prompt}.

\begin{deluxetable}{lcc}
	\tablecaption{Best-fit parameter values for the unbroken power-law model. $\Phi_\mathrm{astro}$ is the normalization of the astrophysical neutrino flux at $100\,\mathrm{TeV}$ and is given in units of $10^{-18}\,\mathrm{GeV^{-1}\,s^{-1}\,sr^{-1}\,cm^{-2}}$. $\Phi_\mathrm{prompt}$ is given in units of the model in \cite{Prompt:ERS}. The normalizations correspond to the sum of neutrinos and antineutrinos.\label{tab:results_physics}}
	\tablehead{\colhead{Parameter} & \colhead{Best-Fit} & \colhead{$68\%\,\mathrm{C.L.}$}}
	\startdata
	  $\Phi_\mathrm{astro}$ & $0.90$ & $0.62-1.20$ \\
      $\gamma_\mathrm{astro}$ & $2.13$ & $2.00-2.26$ \\
      $\Phi_\mathrm{prompt}$ & $0.00$ & $0.00-0.19$ \\
	\enddata
\end{deluxetable}

\begin{figure}
	\centering
	\includegraphics[width=\columnwidth]{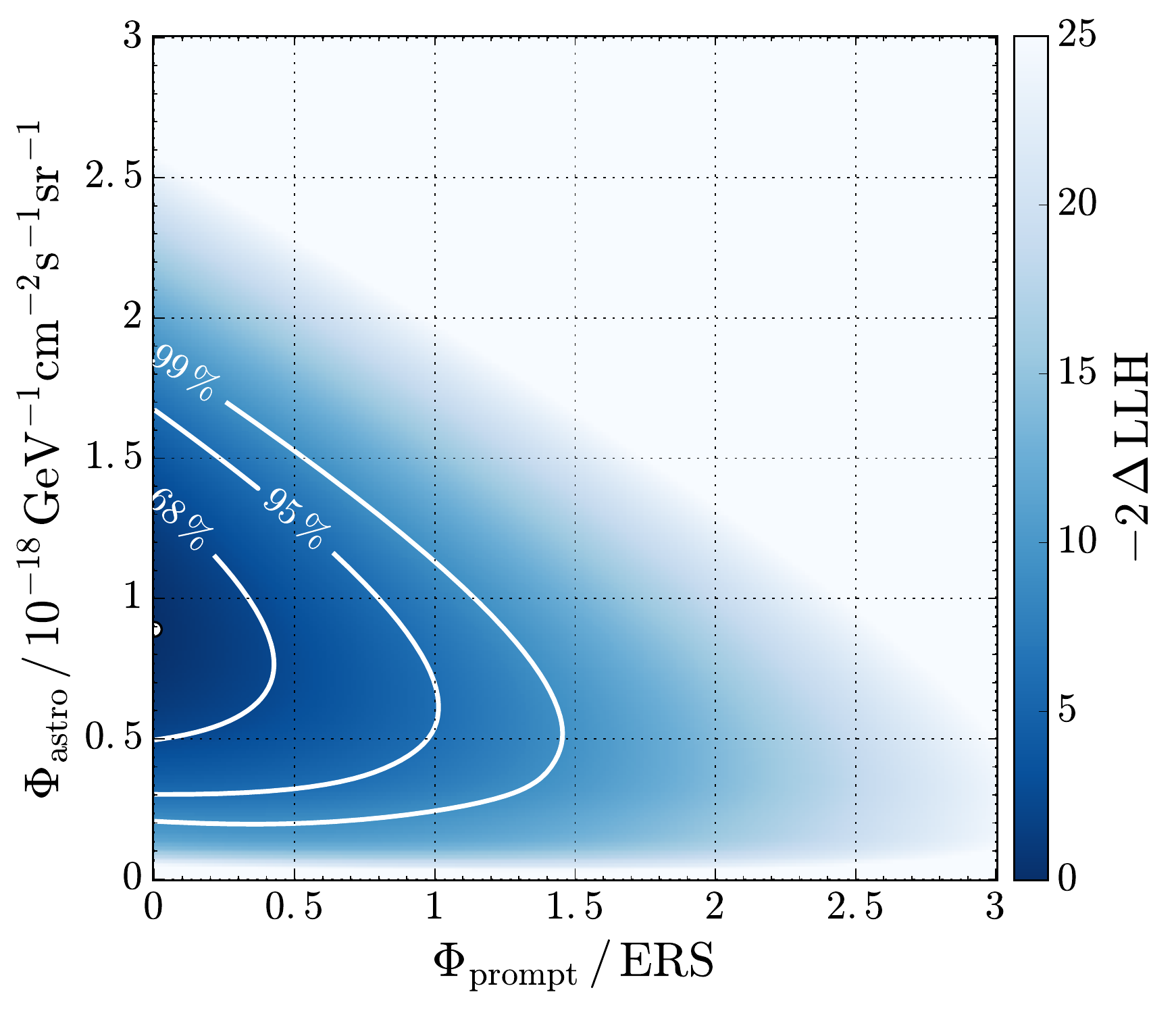}
	\includegraphics[width=\columnwidth]{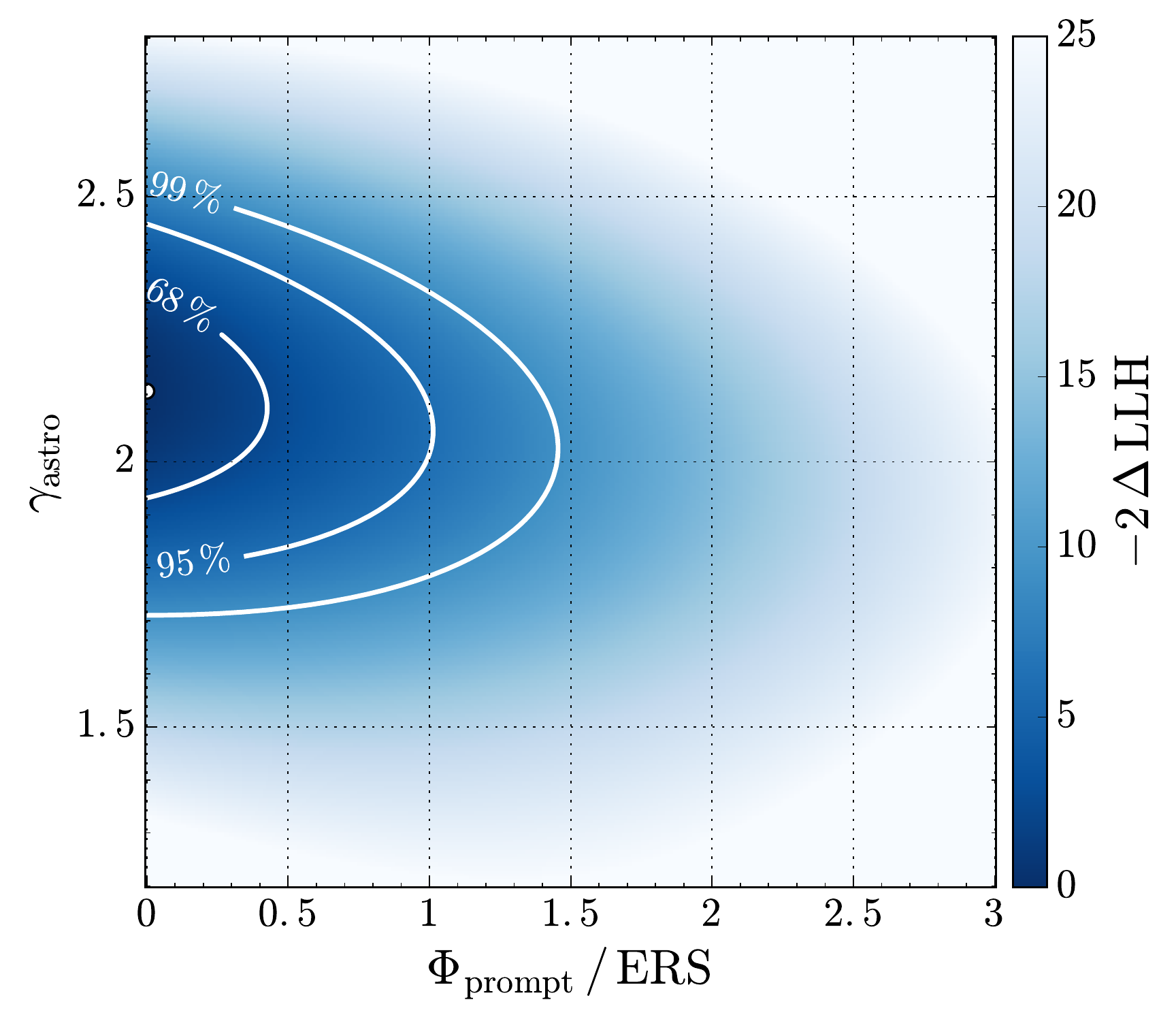}
	\includegraphics[width=\columnwidth]{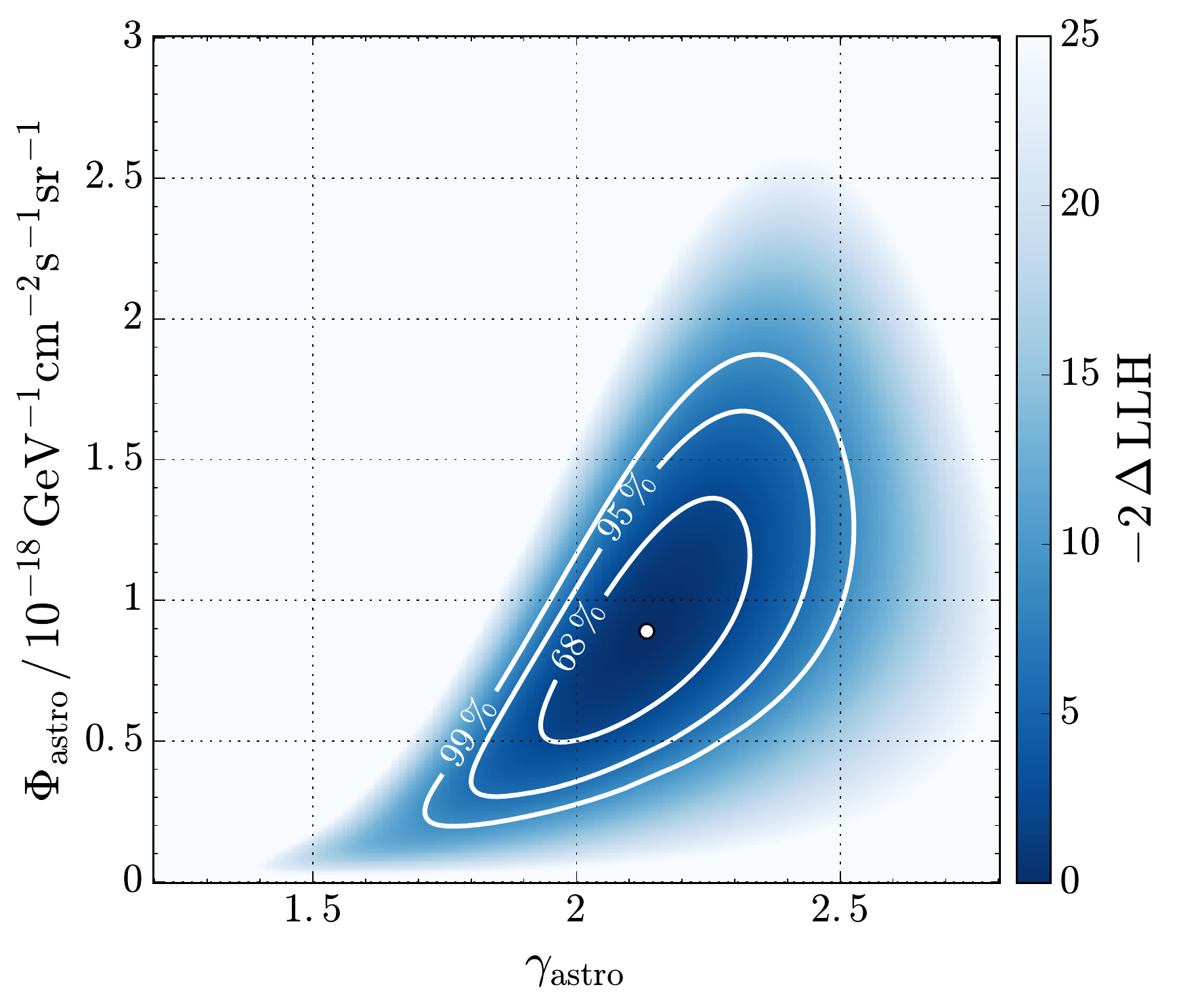}
	\caption{Two-dimensional profile likelihood scans of the astrophysical parameter $\Phi_\mathrm{astro}$, $\gamma_\mathrm{astro}$ and the prompt normalization $\Phi_\mathrm{prompt}$ in units of the model in \cite{Prompt:ERS}. The contours at $68\%$, $95\%$ and $99\%$ CL assuming Wilks' theorem are shown. \label{fig:2dscans}}
\end{figure}

\begin{figure*}
	\centering
	\includegraphics[width=1.6\columnwidth]{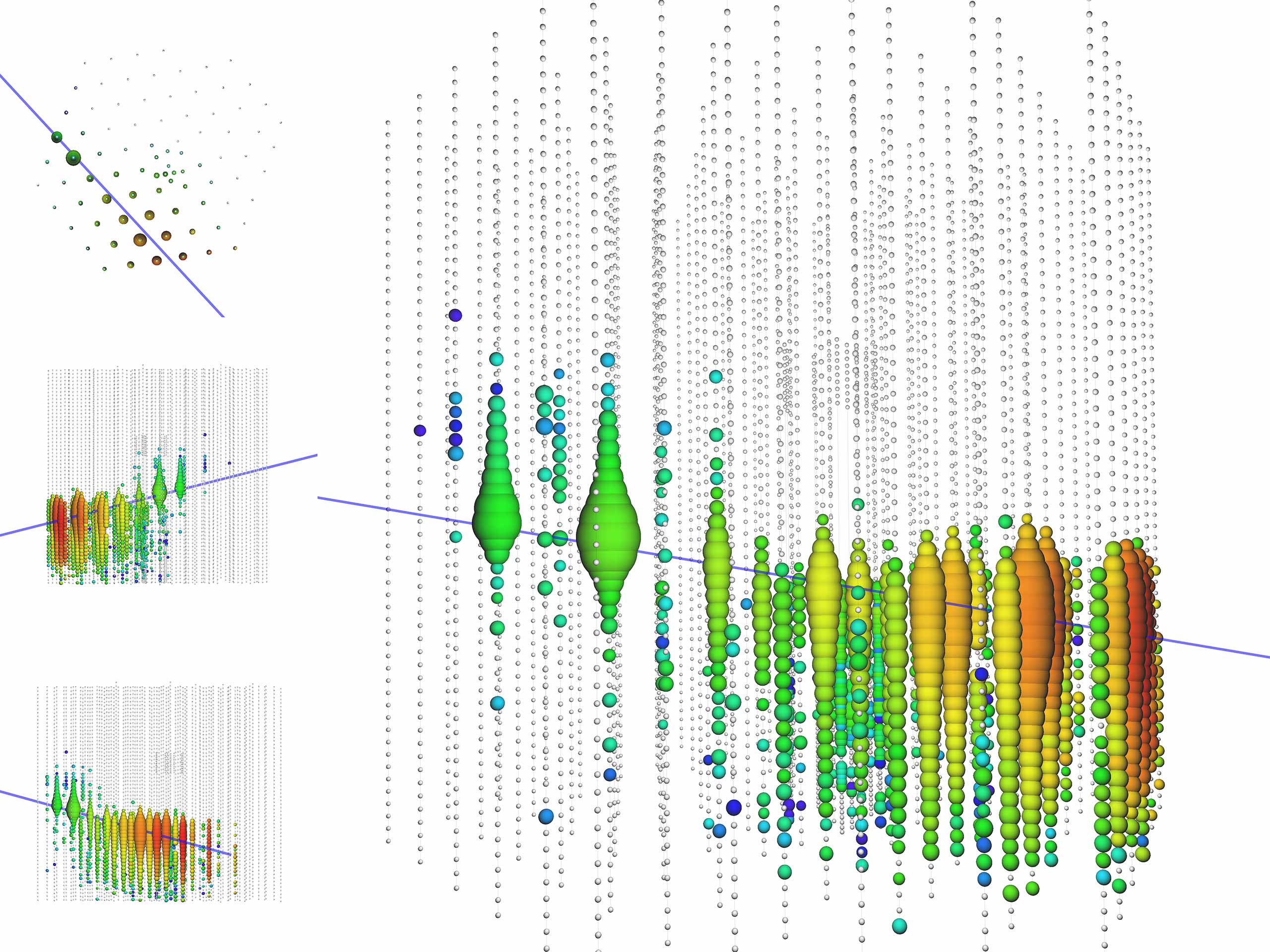}
	\caption{Event view of the PeV track-like event recorded by IceCube on June 11, 2014. Left: Top and two side views. Right: Perspective view. Shown are the IceCube DOMs as black dots. The colors indicate the photon arrival time from red (early) to green (late) and the size of the sphere the amount of measured charge. Note that the scaling is non-linear and a doubling in sphere size corresponds to one hundred times the measured charge. The blue line shows the reconstructed particle track. The reconstructed equatorial coordinates of this event are $\mathrm{dec}=11.42^\circ$ and $\mathrm{ra}=110.63^\circ$. This event deposited an energy of $2.6\pm0.3\,\mathrm{PeV}$ within the detection volume. \label{fig:eventview_PeVtrack}}
\end{figure*}

The two-dimensional contours of the profile likelihood as a function of the signal parameters are shown in Fig. \ref{fig:2dscans}.
While the fitted astrophysical flux normalization is strongly correlated with the astrophysical spectral index, these astrophysical signal parameters are found to be largely independent of the 
prompt flux normalization.

The model assumes an unbroken power-law for the astrophysical signal.
We estimate that neutrinos in the experimental data sample
with energies mainly between $194\,\mathrm{TeV}$ and $7.8\,\mathrm{PeV}$ contribute to this observation.
This energy range is shown in Fig. \ref{fig:bestfit_fluxes}. It defines the central range of neutrino energies that contribute 90\% to the total observed likelihood ratio between the best-fit and the conventional atmospheric-only hypothesis. Note that this definition is different from \cite{IceCube:IC79NuMuDiffuse, IceCube:MESE2Years}.

\subsection{Multi-PeV track-like event} \label{sec:results_PeVTrack}
The selected data include one exceptionally high-energy muon event 
that is shown in Fig. \ref{fig:eventview_PeVtrack} \citep{schoenen2015detection}.
The deposited energy has been measured to $(2.6\pm 0.3)$\,PeV
of equivalent electromagnetic energy \cite{Aartsen:2013vja}. 
Assuming the best-fit atmospheric energy spectrum from this analysis (see Fig. \ref{fig:bestfit_fluxes}) 
the p-value of this event being of atmospheric origin
has been estimated to be less than $0.005\%$, strongly suggesting an astrophysical origin. 

The segmented energy loss reconstruction described in \cite{Aartsen:2013vja} can be used to reconstruct the direction of through-going muons. This includes the timing of not only the first photon but all photons as well as the total number of photons.
The reconstructed direction of the event is given in Tab. \ref{tab:he:events} 
and discussed in Sec. \ref{sec:aniso:kloppo}.

In order to estimate the angular uncertainty and the 
most likely muon and neutrino energy we have simulated events with energies according to our best-fit energy spectrum with directions varying by $1^\circ$ around the best-fit direction. Additionally, the position where the muon enters the instrumented volume has been varied within $10\,\mathrm{m}$.
Systematic uncertainties due to the lack of knowledge about the optical ice properties are taken into account by varying the ice model parameters within their uncertainties during the simulation.

Based on these simulations we evaluate the muon energy at the point of entrance into the instrumented volume, that results in the observed deposited energy. The obtained median muon energy is $(4.5\pm1.2)\,\unit{PeV}$ where the error
range corresponds to 68\% C.L.

\begin{figure}
	\includegraphics[width=1.\columnwidth]{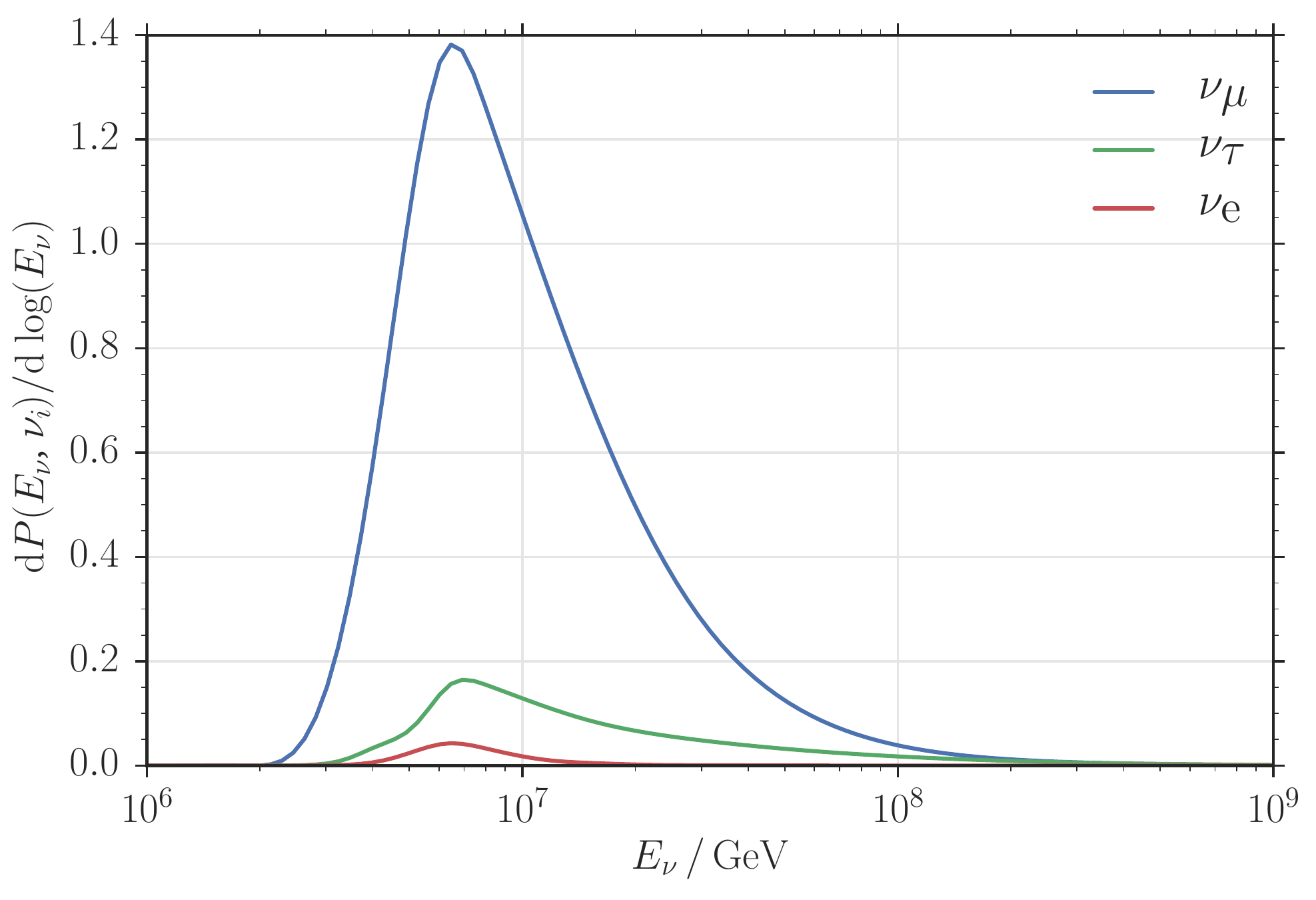}
	\caption{Probability distribution of primary neutrino energies that could result in the observed multi-PeV track-like event. The total probabilities for the different flavors are $87.7\%$, $10.9\%$ and $1.4\%$ for $\nu_\mu$, $\nu_\tau$ and $\bar{\nu}_e $, respectively.\label{fig:flux:kloppoen}}
\end{figure}

For the estimation of the median expected neutrino energy we have taken into account that high energy muons arise not only from $\nu_\mu $ charged current interactions but also from muonic decay of charged current $\nu_\tau $ interactions and muonic $W^- $ decays in $\bar{\nu}_e + e^- \to W^- $ interactions.
Here, we assume the best-fit astrophysical spectrum and an equal flux of all flavors but include the effects of the Earth's absorption for the
 specific declination of the event. Under these assumptions, we find 
$ 87.7\% $ probability of a primary $\nu_\mu $, $10.9\% $ for a primary 
$\nu_\tau $ and $ 1.4\%$ for a primary $\bar{\nu}_e $.
The respective probability distributions of primary neutrino energy are shown in Fig. \ref{fig:flux:kloppoen}.
The expected neutrino energy depends on the primary flavor.
The median expected muon neutrino energy is $8.7\,\mathrm{PeV}$ for the above assumptions.

\begin{figure}
	\centering
	\includegraphics[width=1.\columnwidth]{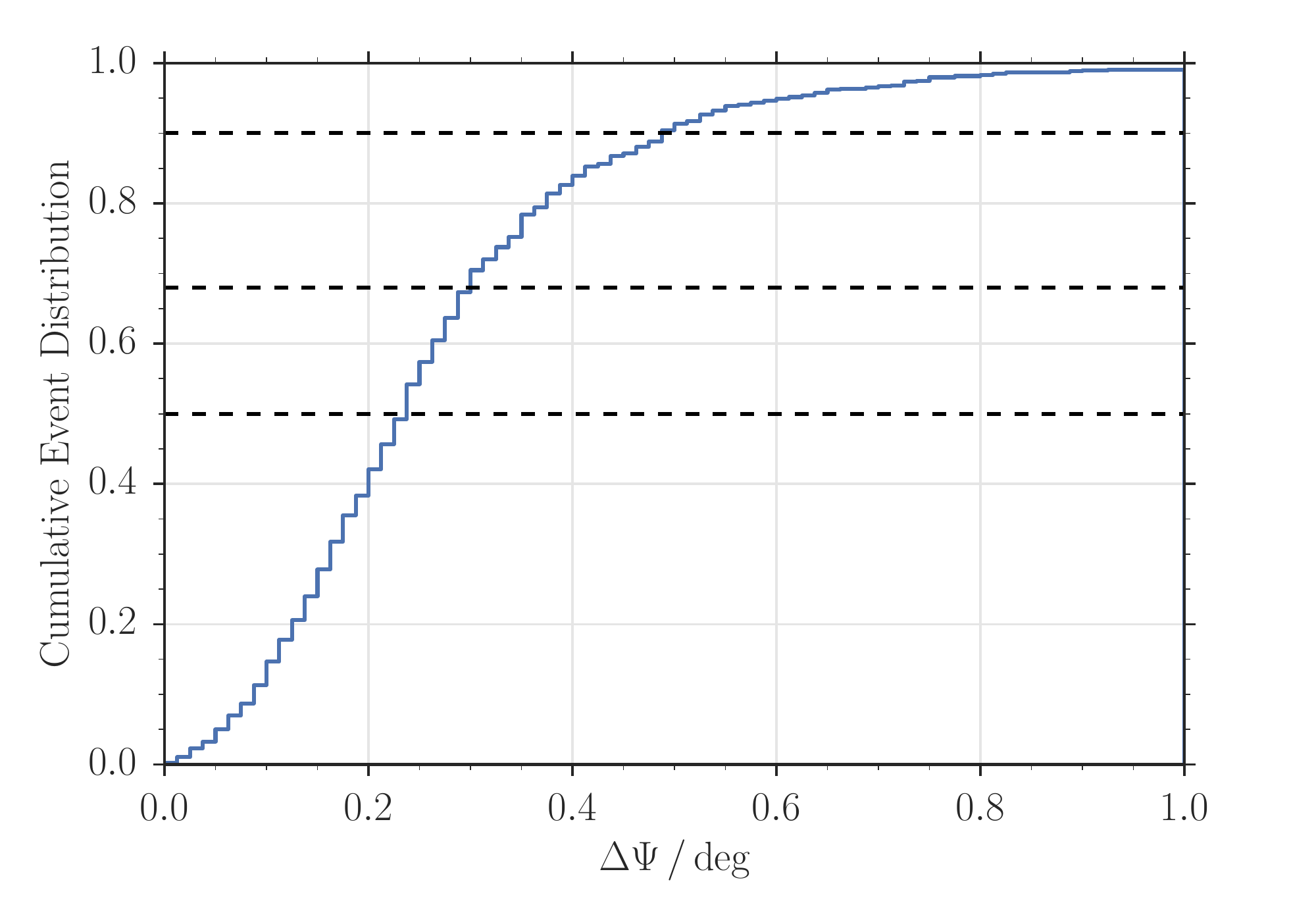}
	\caption{Angular reconstruction uncertainty of the multi-PeV track-like event. The estimate is based on an ensemble of simulated events with similar deposited energy, direction and entry point into the fiducial volume. The simulation takes into account the ice uncertainties (cf. Sec. \ref{sec:systematics}). Including statistical and systematic uncertainties, $50\%$ ($99\%$) of the events are reconstructed better than $0.23^\circ$ ($0.90^\circ$). \label{fig:angular_resolution_PeVtrack}}
\end{figure}

The angular reconstruction uncertainty including systematic uncertainties of the Antarctic ice (cf. Sec. \ref{sec:systematics}) can be estimated from the aforementioned dedicated simulation. Figure \ref{fig:angular_resolution_PeVtrack} shows the angular reconstruction uncertainty for an ensemble of events with similar deposited energy, direction and entry point into the fiducial volume. The angular reconstruction uncertainty is given by the angular distance between the true and the reconstructed muon direction. The median angular uncertainty is $0.23^{\circ}$ and the 99\% containment is $0.9^\circ$.
Details of the studies of the multi-PeV track-like event are shown in \cite{PhDThesis:Raedel}.

\subsection{Test for a spectral cut-off}
\begin{figure}
	\centering
	\includegraphics[width=\columnwidth]{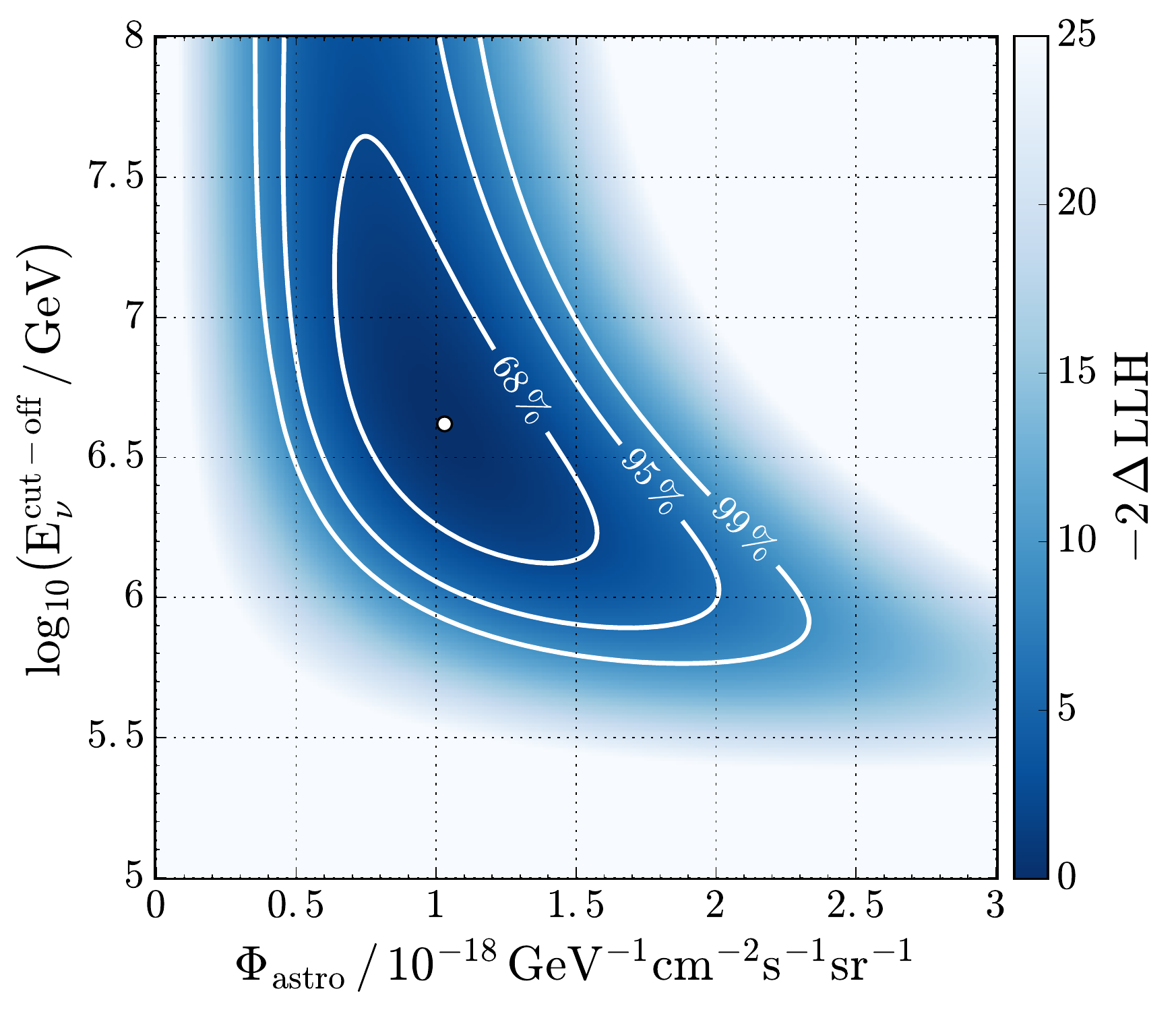}
	\includegraphics[width=\columnwidth]{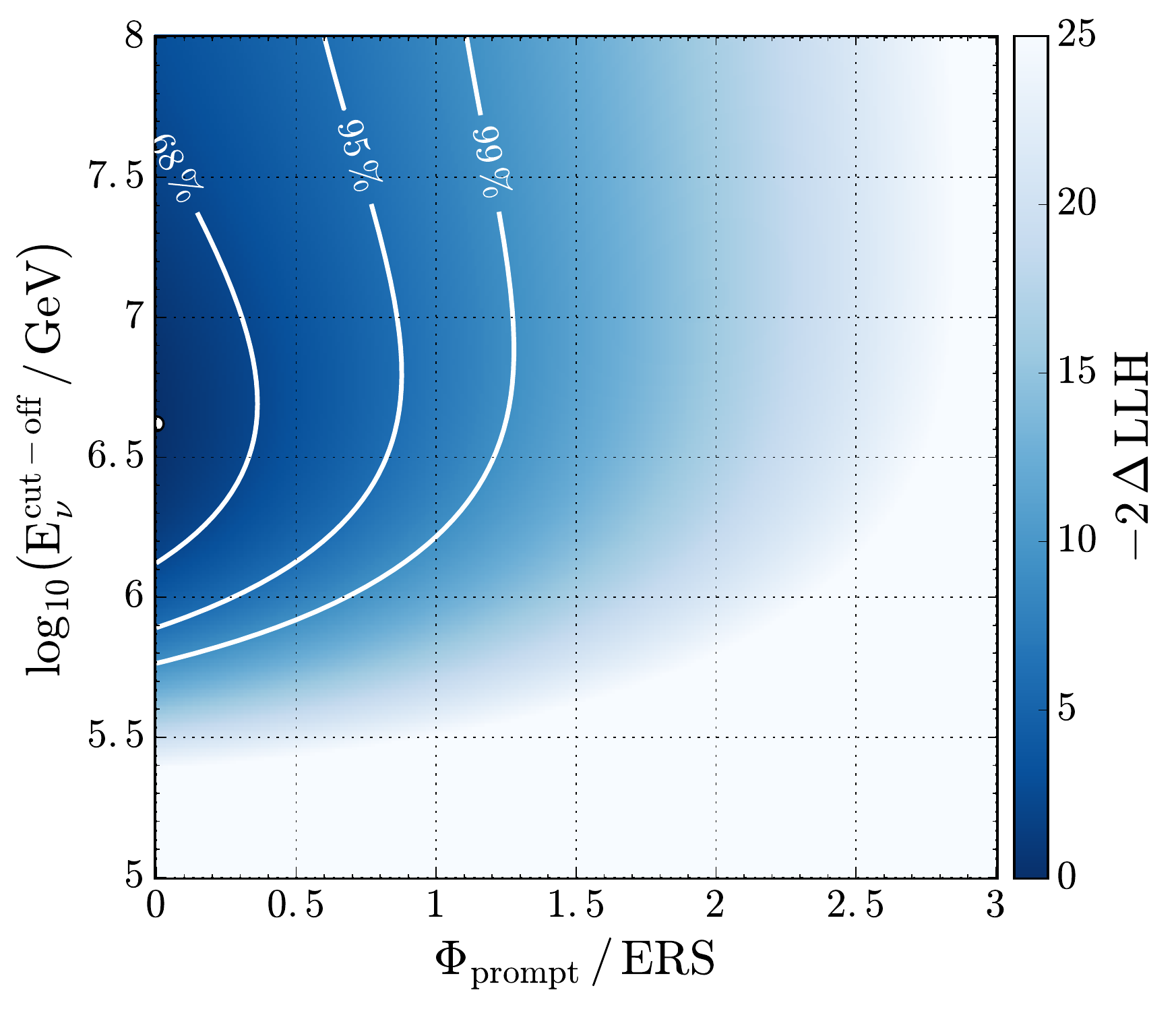}
	\includegraphics[width=\columnwidth]{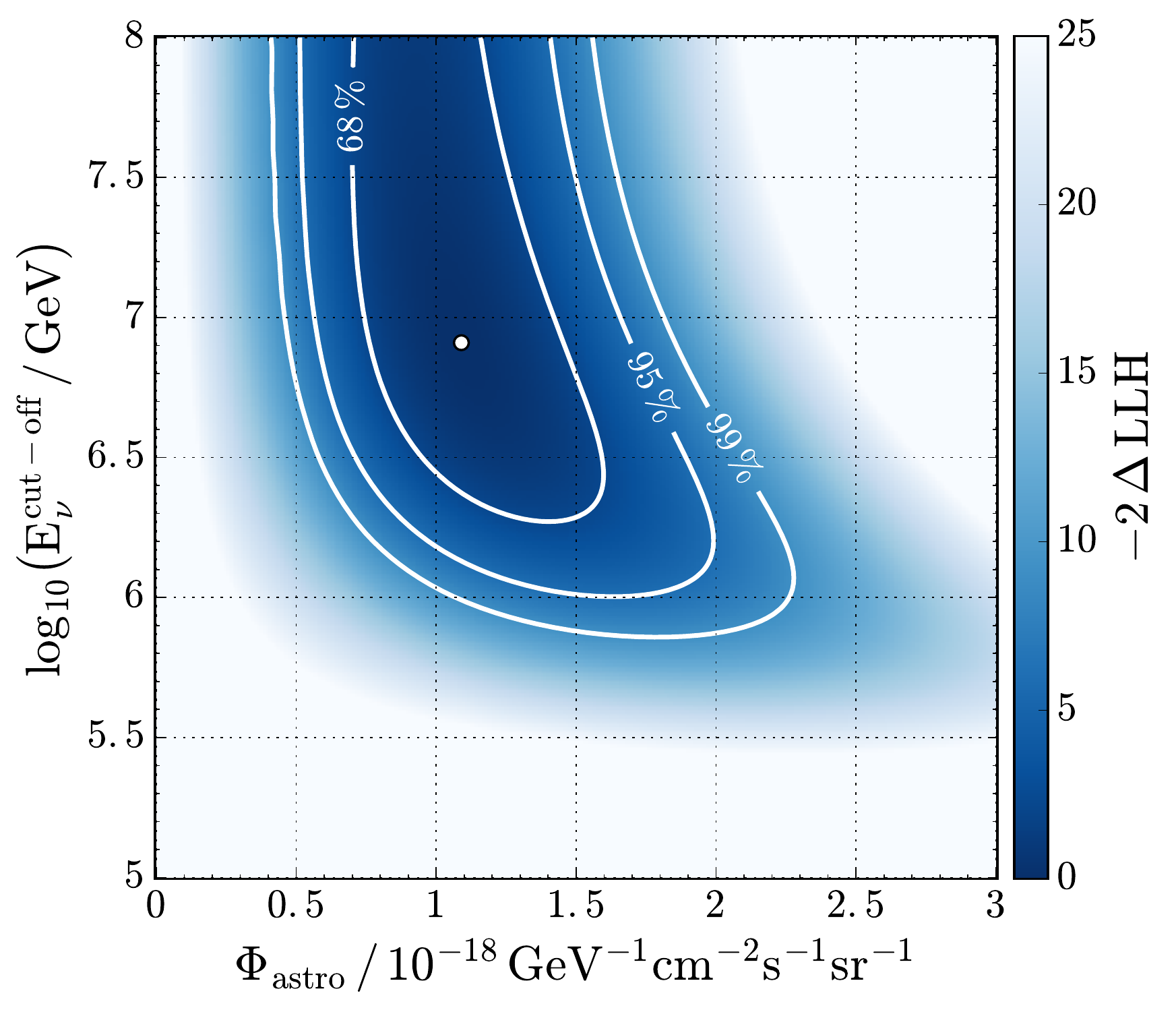}
	\caption{Two-dimensional profile likelihood scans of the astrophysical parameters $\Phi_\mathrm{astro}$, $E_\nu^\mathrm{cut-off}$ and the prompt normalization $\Phi_\mathrm{prompt}$ in units of the model in \cite{Prompt:ERS}. The contour lines at $68\%$, $95\%$ and $99\%$ CL assume Wilks' theorem. For the top and middle figure the spectral index is fixed to $\gamma_\mathrm{astro}=2$, while 
in the bottom figure it is fixed to the best-fit value $\gamma_\mathrm{astro}=2.13$. The white dots indicate the best-fit values. \label{fig:2dscans_cutoff}}
\end{figure}

The default hypothesis of an unbroken power-law
is tested against the hypothesis of a spectral cut-off. For this, an exponential energy cut-off $E_\nu^\mathrm{cut-off}$ is added to the astrophysical neutrino flux:
\begin{equation}
  \Phi_{\nu+\overline{\nu}} = \Phi_\mathrm{astro} \cdot \exp\left(-\frac{E_\nu}{E_\nu^\mathrm{cut-off}}\right) \cdot \left(\frac{E_\nu}{100\,\mathrm{TeV}}\right)^\mathrm{-\gamma_\mathrm{astro}}.
\end{equation}
In the fit the spectral index $\gamma_\mathrm{astro}$ is highly degenerate with an exponential energy cut-off $E_\nu^\mathrm{cut-off}$, therefore two scenarios with fixed spectral indices have been tested.
For the spectral indices the benchmark model with $\gamma_\mathrm{astro}=2$ and the best-fit value $\gamma_\mathrm{astro}=2.13$ are chosen. Figure \ref{fig:2dscans_cutoff} shows the two-dimensional contours of the profile likelihood as a function of the signal parameters $\Phi_\mathrm{astro}$, $E_\nu^\mathrm{cut-off}$ and $\Phi_\mathrm{prompt}$. For the benchmark model a cut-off is slightly preferred at the level of one standard deviation.
This is an expected behavior as the actual best-fit spectral index 
is softer. Thus, fixing the spectral index to a harder spectrum will result in a slight deficit at the highest neutrino energies. When fixing the spectral index to the best-fit value for an unbroken power law, this slight preference for an exponential cut-off disappears. These results are nearly independent of the prompt flux normalization.

\subsection{Unfolded astrophysical spectrum}
\begin{figure}
	\centering
	\includegraphics[width=1.\columnwidth]{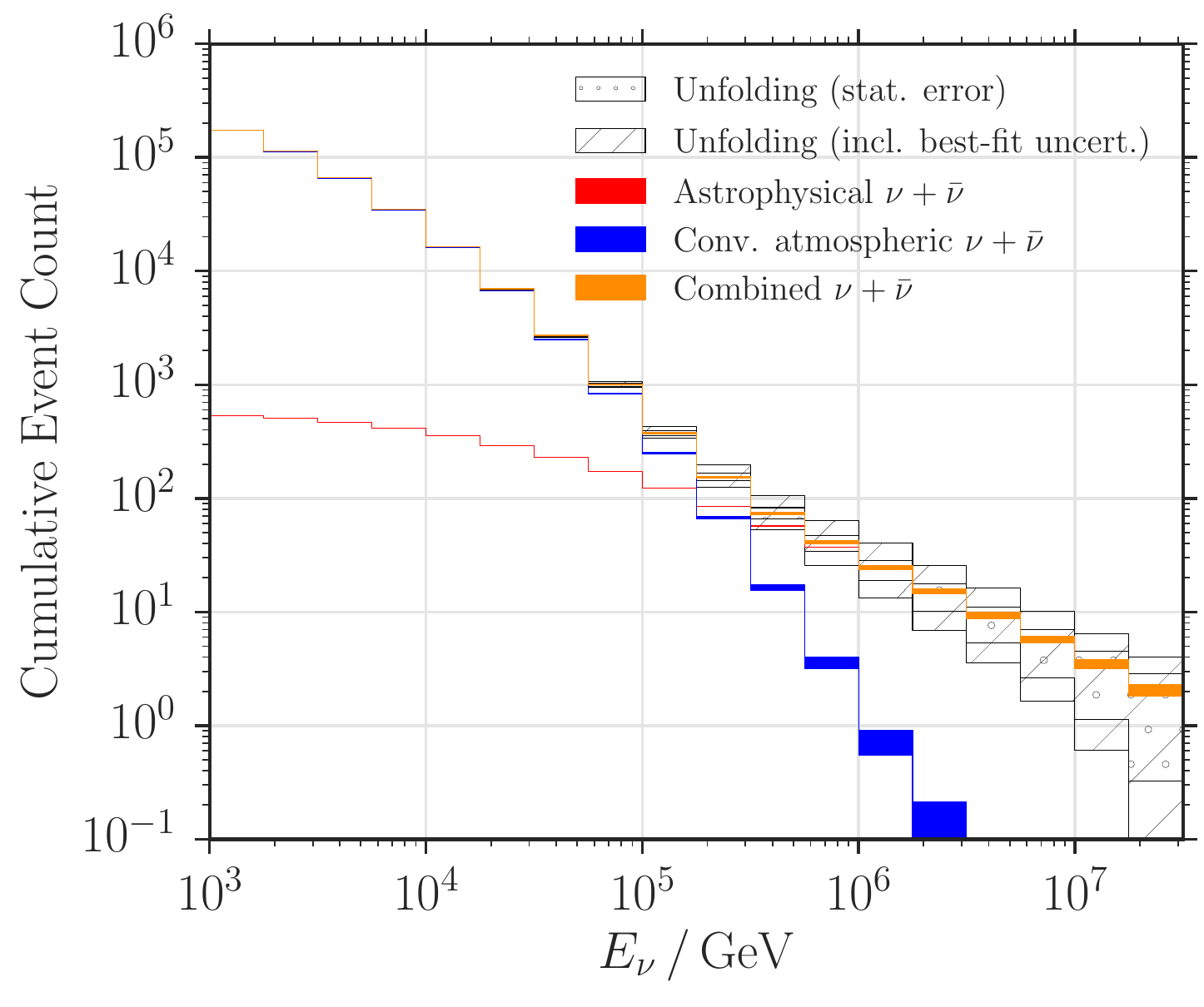}
	\caption{Cumulative distribution of the parametric unfolded neutrino energy spectrum for the 6 year data sample assuming the best-fit spectrum as given by this analysis. Blue / red corresponds to the conventional atmospheric / astrophysical expectation weighted to the best-fit spectrum. Orange represents the sum of both expectations. The parametric unfolded data is shown as hatched band where the gray band shows the statistical uncertainty and the white band additionally the effect of the uncertainties on the fitted astrophysical flux parameters. \label{fig:numu_energy_unfold}}
\end{figure}

\begin{figure}
	\centering
	\includegraphics[width=1.\columnwidth]{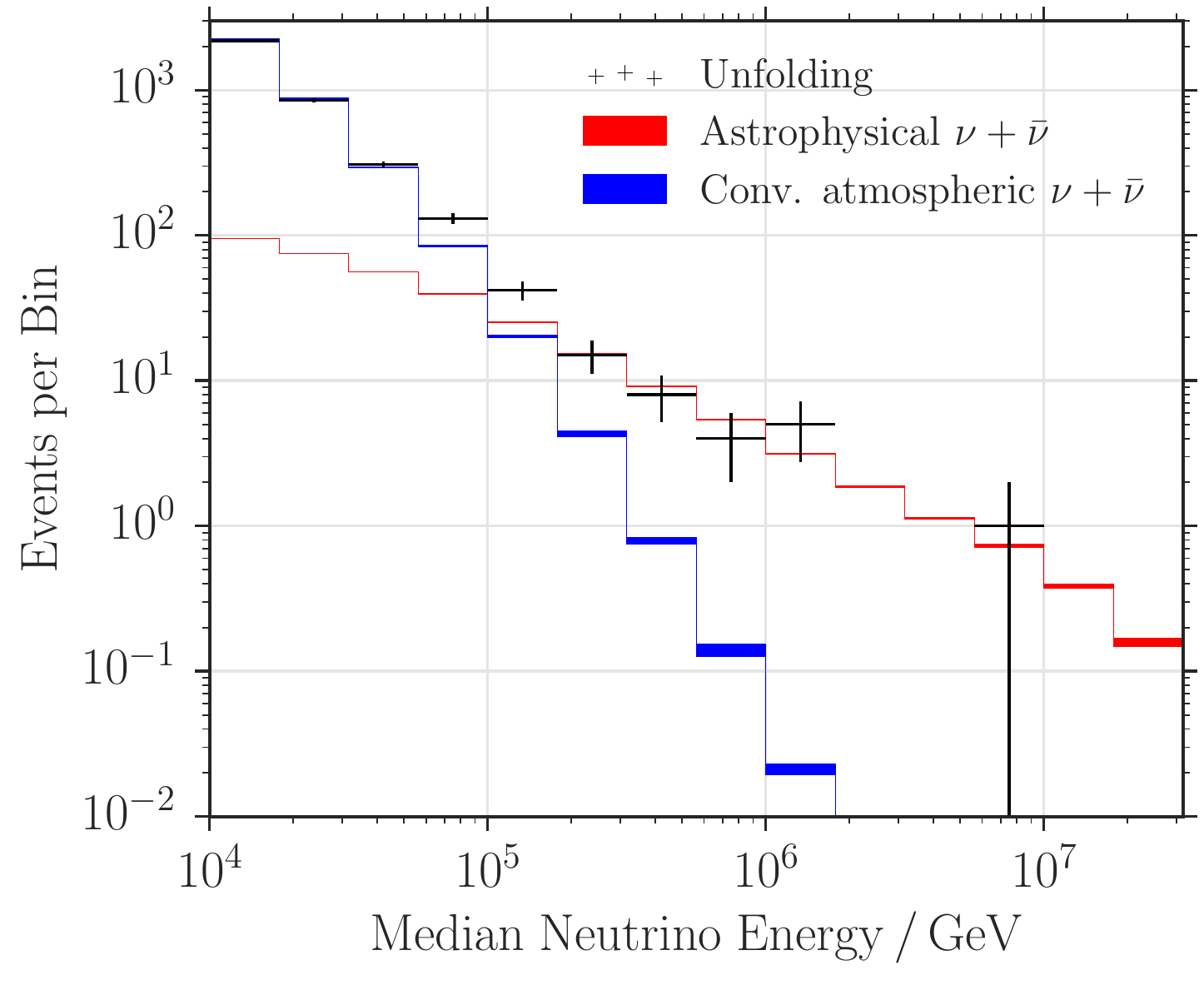}
	\caption{Distribution of the median expected neutrino energy assuming the best-fit spectrum as given by this analysis. The black crosses corresponds to experimental data and blue / red to the conventional atmospheric / astrophysical expectation weighted to the best-fit spectrum. \label{fig:median_numu_energy}}
\end{figure}

The best-fit results for the neutrino energy spectrum as quoted in Tab. \ref{tab:results_physics} and the knowledge about the connection between the reconstructed muon and true neutrino energy can be used to unfold a neutrino energy distribution for the six years sample (cf. Sec. \ref{sec:analysis_method:unfolding}). The results of this parametric unfolding are shown in Fig. \ref{fig:numu_energy_unfold}.
as cumulative energy distribution of the number of neutrinos with energies greater than $E_\nu$. The statistical error band is given by the square root of this number. The error band that corresponds to the uncertainty on the astrophysical flux is determined by varying the astrophysical spectrum within the measured uncertainties on the astrophysical flux parameters. Based on the per-event probability density function $P(E_\nu | E^i_\mathrm{reco})$ also the median neutrino energy for each event can be calculated. Figure \ref{fig:median_numu_energy} shows the distribution of the median neutrino energies for the six year sample.

In both distributions a clear excess above approximately $100\,\mathrm{TeV}$ in neutrino energy is visible, and is not compatible with the atmospheric background expectation.
Although only a single event with energy greater than a PeV has been observed, we can infer from our fit and from the relation between muon energy and energy of the parent neutrino that there are most likely several neutrinos with energies above a PeV in the 6 year sample.
 
\subsection{Discussion}\label{sec:flux:discuss}
\begin{figure}
	\centering
	\includegraphics[width=1.\columnwidth]{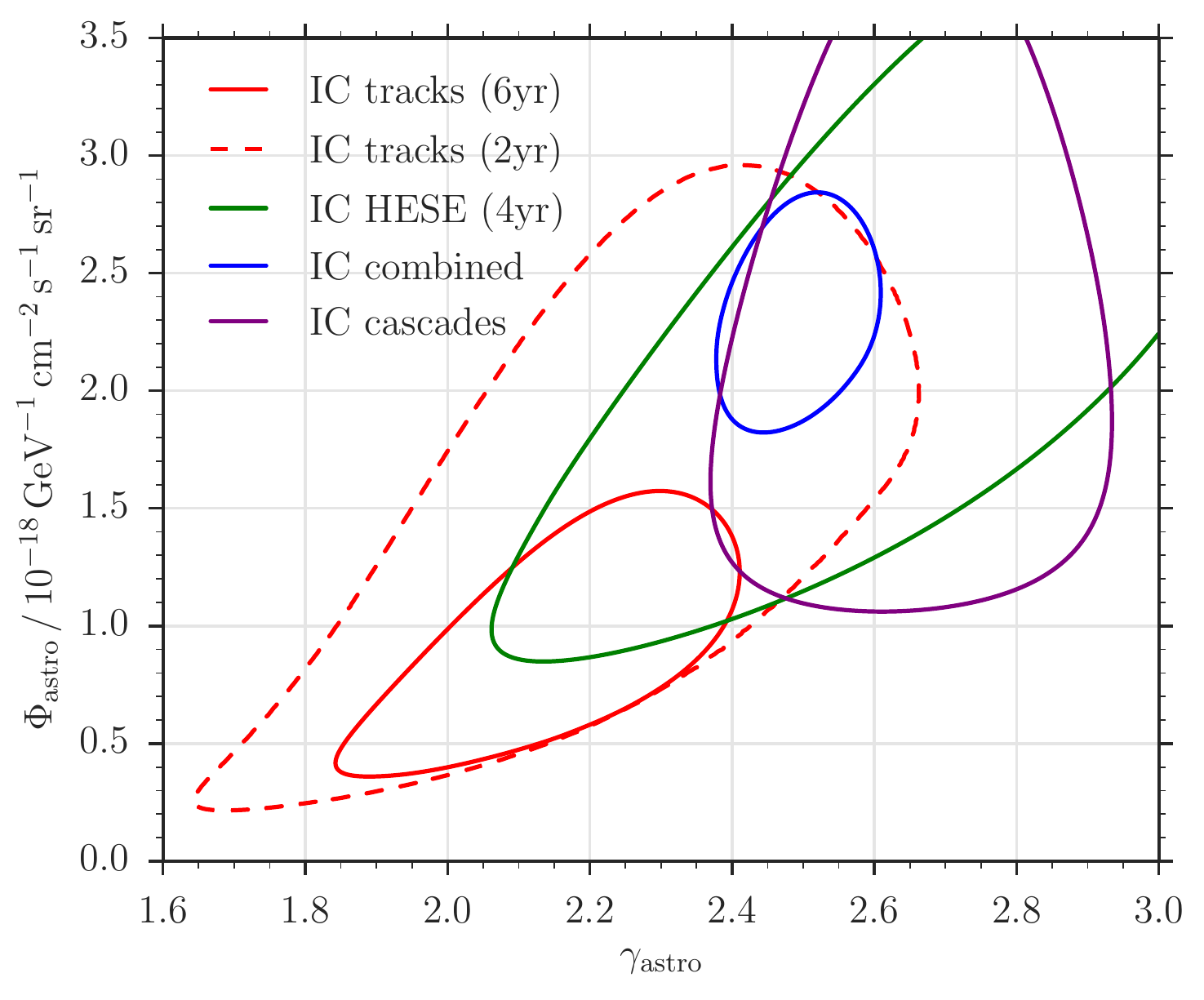}
	\caption{Results of different IceCube analyses measuring the astrophysical flux parameters $\Phi_\mathrm{astro}$ and $\gamma_\mathrm{astro}$. The contour lines show the $90\%$ CL. The result of this analysis (IC tracks, 6yr) is shown by the red solid contour line. The contour obtained by the previous measurement using through-going muons \citep{IceCube:IC79NuMuDiffuse} (IC tracks, 2yr) is the red dashed line. In addition, the results for the most recent analysis of starting events \citep{Aartsen:2015zva:hese} (IC HESE, 4yr), the complementary cascade channel \citep{Aartsen:2015zva:casc} (IC cascades) and an analysis combining different IceCube results \citep{Aartsen:2015zva:GlobalFit} (IC combined) are shown. The result of this analysis (red, solid) and IC combined are incompatible at $3.3\sigma$ (two-sided significance). \label{fig:2dcontour_IceCube_comparison}}
\end{figure}

This analysis found an astrophysical spectral index of $\gamma = 2.13 \pm 0.13$, that is harder than previously reported measurements, see e.g. \cite{Aartsen:2015zva:hese}, \cite{Aartsen:2015zva:casc} and \cite{Aartsen:2015zva:GlobalFit}.
We refer to these analyses in the rest of the section as starting event analysis, cascade analysis and combined analysis, respectively.
Figure \ref{fig:2dcontour_IceCube_comparison} compares the measured astrophysical neutrino flux normalization and spectral index with these results and the previous measurement using through-going muons \cite{IceCube:IC79NuMuDiffuse}. 

While the sample used in the cascade analysis is completely statistically 
independent, the starting event analysis and global fit have an overlap in events.
The combined fit includes three years of muon data from 2009-2012 based on \cite{IceCube:IC59NuMuDiffuse,IceCube:IC79NuMuDiffuse}.
The starting event analysis includes a small fraction (6\%) of up-going 
muons that start within the detector, that are also included here.
However, these three analyses are strongly dominated by independent 
cascade-like events of which a large fraction originates from the 
Southern hemisphere. For the starting event analysis 73\% of the events above 100\,TeV are down-going and 93\% of these are cascade-like.
For the investigation of the tension in the observed energy spectrum of astrophysical neutrinos, the assumption of statistical independence is reasonably well justified but will result in a lower limit on the tension.

The combined analysis finds the smallest confidence region of the three aforementioned results.
The p-value for obtaining the combined fit result and the result reported here from an unbroken powerlaw flux is $3.3\sigma$, and is therefore in significant tension.
For the discussion, it is important to highlight the systematic differences 
between these measurements. The threshold for the up-going muon signal
 is a few hundred TeV while 
astrophysical starting events are detected above a few times $10\,\mathrm{TeV}$.
It should be noted that for the overlapping energy region $>200 $\,TeV
 the measured fluxes for the cascade dominated channels
are in good agreement with the results reported here, as shown in
Fig. \ref{fig:bestfit_fluxes}. As a conclusion, we confirm
for the Northern hemisphere a flux of muon neutrinos that is
generally consistent with the observed all flavor flux in the Southern hemisphere,
but which is in tension with the assumption of a single power law describing this and previous observations with a lower energy threshold at the same time.

It is expected that for a galactic origin the neutrino flux should be correlated with the galactic plane. 
It is generally assumed that the contribution from the galactic plane and galactic sources is stronger in the Southern hemisphere, which e.g. includes the galactic Center.
The measured astrophysical flux is not strongly affected by a split in right ascension (see Sec.~\ref{sec:aniso:galactic:plane}), where one region includes the part of the galactic plane which is visible in the Northern sky and the other does not. This can be interpreted as an indication
that the flux observed here is mostly of extra-galactic origin. 

The observed tension may arise either from a spectral break at
 lower energies for the same sources or from 
an additional flux component, e.g. expected from galactic sources or the galactic plane, that is sub-dominant at the high energies to which 
this analysis is sensitive.

\begin{figure}
	\centering
	\includegraphics[width=1.0\columnwidth]{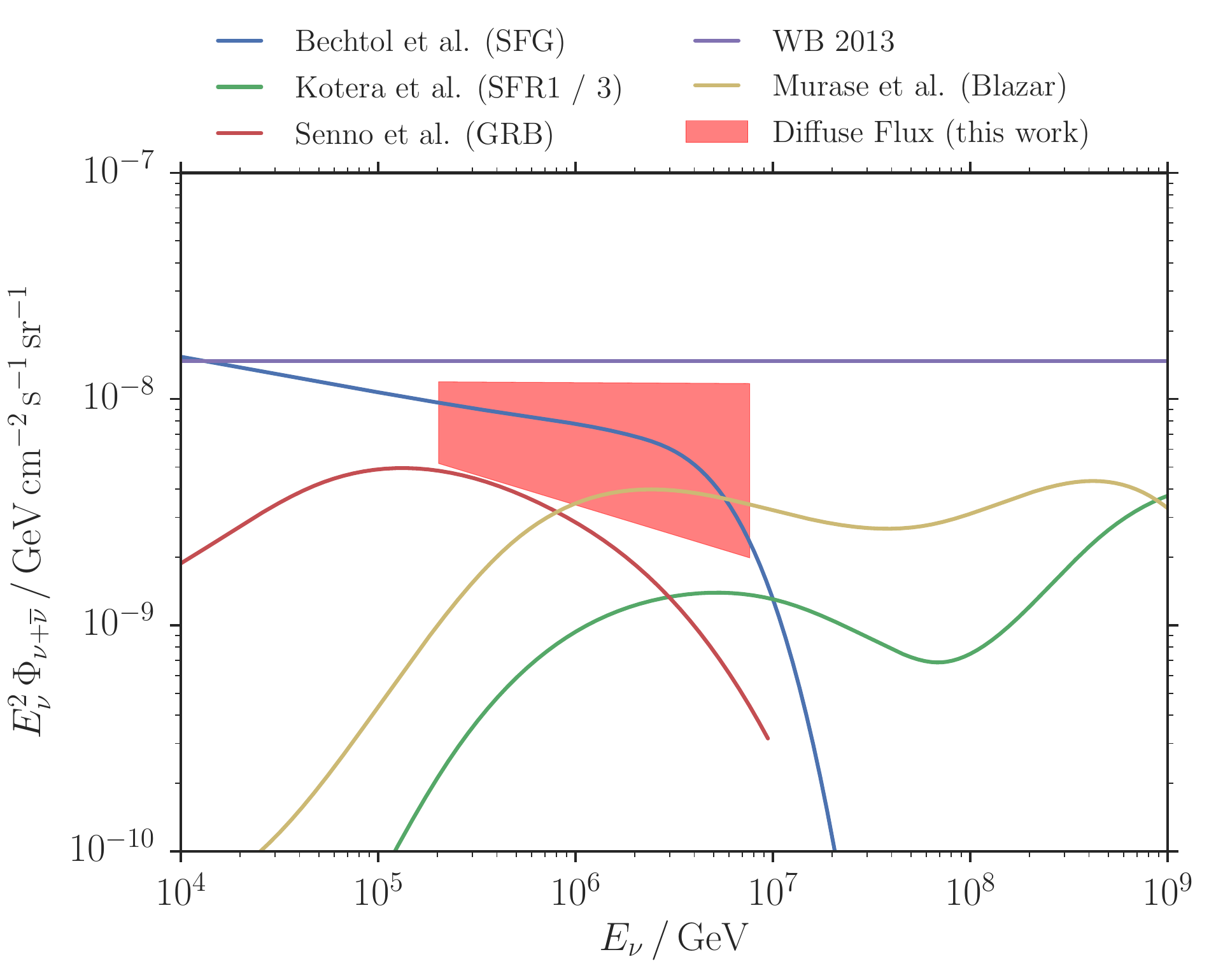}
	\caption{Comparison of the measured diffuse astrophysical muon neutrino flux (cf. Fig. \ref{fig:bestfit_fluxes}) with theoretical neutrino flux predictions corresponding to different source types \citep{Kotera:2010, MuraseBlazers:2014, Bechtol:2015, Senno:2015}. Since \cite{MuraseBlazers:2014} predicts a lower and upper flux bound for neutrinos originating from Blazars the central line between both bounds is shown. The purple line shows the Waxman-Bahcall upper bound \citep{WaxmanUpperBound:2013}. \label{fig:astro_flux_comparisons}}
\end{figure}

Figure \ref{fig:astro_flux_comparisons} compares the measured diffuse astrophysical muon neutrino flux to theoretical flux predictions corresponding to different source types. The measured flux is within its uncertainties slightly below the Waxman-Bahcall upper bound \citep{WaxmanUpperBound:2013}. \cite{Senno:2015} predict a diffuse neutrino flux originating from gamma-ray burst which is currently not ruled out \citep{Aartsen:2016GRB1, Aartsen:2016GRB2}. A flux of cosmogenic neutrinos as predicted by \cite{Kotera:2010} would only contribute subdominantly to the measured astrophysical neutrino flux. Neutrino fluxes from blazars and star-forming galaxies are predicted by e.g. \cite{MuraseBlazers:2014} and \cite{Bechtol:2015}, respectively. \cite{Glusenkamp:2015jca} already constrains this blazar model. These fluxes are of the same order of magnitude as the measured flux within the given uncertainty band. However, due to the small statistics at high energies we cannot differentiate if the measured astrophysical neutrino flux corresponds to a neutrino flux originating from a specific source type or if it is a combination of different source types.

\section{Analysis of arrival directions and search for anisotropies} \label{sec:aniso}

\subsection{Arrival directions of highest energy events} \label{sec:aniso:kloppo}
The multi-PeV event discussed in Sec. \ref{sec:results_PeVTrack} has a high probability of being astrophysical. Therefore, it is particularly interesting to correlate such an event with potential sources. 

\begin{figure}
	\centering
	\includegraphics[width=1.\columnwidth]{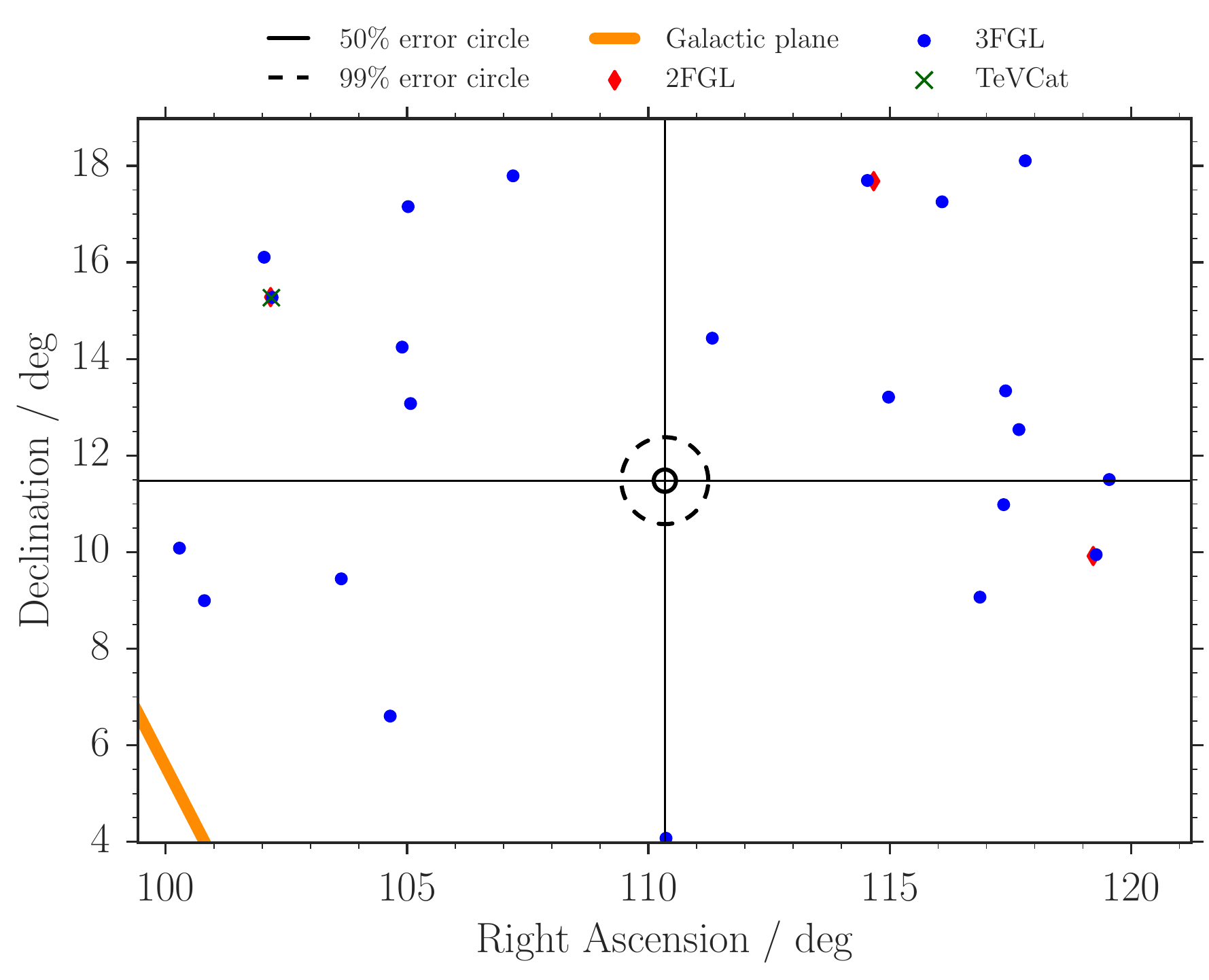}
	\caption{Window centered around the arrival direction of the multi-PeV track-like event. The solid (dashed) black line shows the $50\%$ ($99\%$) error circle for the angular reconstruction. The orange line indicates the galactic plane. Additionally, the gamma-ray sources of the catalogs \cite{TeVCat,Fermi3FGL,Fermi2FGL} within the window are shown. \label{fig:skymap_PeVtrack}}
\end{figure}

Figure \ref{fig:skymap_PeVtrack} shows the direction of the event with its angular uncertainty and nearby high-energy gamma-ray sources from \cite{Fermi2FGL, Fermi3FGL, TeVCat} 
in a window centered around the arrival direction. 
The closest source is multiple degrees away which is much larger than the angular error estimate.

\begin{figure*}
	\centering
	\includegraphics[width=0.8\textwidth]{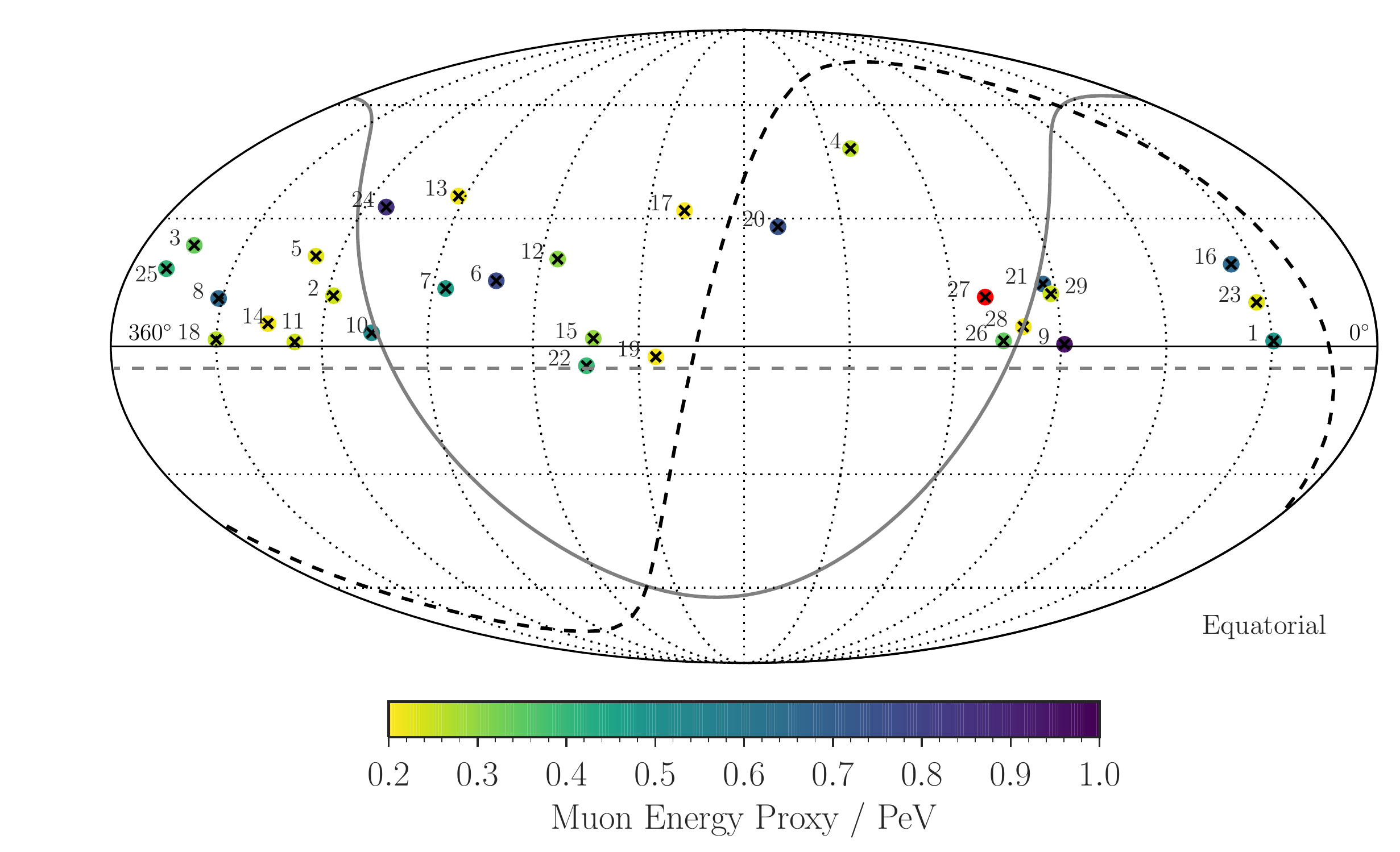}
	\caption{Arrival directions of events with a muon energy proxy above 200TeV. Given the best-fit spectrum the ratio of astrophysical to atmospheric events is about two to one. The horizontal dashed gray line shows the applied zenith angle cut of $85^\circ$. The curved gray line indicates the galactic plane and the dashed black line the supergalactic plane \citep{ISI:000085634700018}. The multi-PeV track event is shown as a red dot and the energy proxy value listed in Tab. \ref{tab:he:events}. \label{fig:skymap:high_energy_events}}
\end{figure*}

For events that have a muon energy proxy above $200\,\mathrm{TeV}$ we expect roughly twice as many events with an astrophysical origin than with an atmospheric origin, assuming the best-fit spectrum. 
Figure \ref{fig:skymap:high_energy_events} shows the arrival direction of these events. Most events are located relatively close to the horizon where the Earth is not yet opaque to high energy neutrinos. Table \ref{tab:he:events} summarizes the per-event information. No obvious correlation with gamma-ray
sources in the following catalogs \cite{Fermi2FGL, Fermi3FGL, TeVCat} were found. However, event 10 is close to the extended TeV source HESS J1857+026 \citep{TeVCat}.

A dedicated analysis searching for clusters in the neutrino arrival directions has been performed and found no evidence for a neutrino point source \citep{IceCube:PS7yr}.

\begin{deluxetable*}{llllllllll}
	\tablecaption{Summary of highest energy events above $200\,\mathrm{TeV}$ in all years. The horizontal lines separate the different data sets IC59, IC79, IC2011 and IC2012-2014. The signalness is defined as the ratio of the astrophysical expectation over the sum of the atmospheric and astrophysical expectations for a given energy proxy and the best-fit spectrum. The signalness decreases up to about 10\% when taking into account a prompt flux at the conservative upper limit of $1.06\times$ERS (cf. Sec. \ref{sec:prompt}). The angular errors are statistical errors only and do not include systematics. \label{tab:he:events}}
	\tablehead{\colhead{ID} & \colhead{MJD} & \colhead{Signalness} & \colhead{Energy Proxy (TeV)} & \colhead{Decl. (deg)} & \colhead{$50\%\,\mathrm{C.L.}$} & \colhead{$90\%\,\mathrm{C.L.}$} & \colhead{R.A. (deg)} & \colhead{$50\%\,\mathrm{C.L.}$} & \colhead{$90\%\,\mathrm{C.L.}$}}

	\startdata
		1 & 55056.70\tablenotemark{a}    & 0.78 & 480 & 1.23 & ${}_{-0.08}^{+0.08}$ & ${}_{-0.22}^{+0.18}$ & 29.51 & ${}_{-0.17}^{+0.15}$ & ${}_{-0.38}^{+0.40}$  \\
		2 & 55141.13\tablenotemark{a}    & 0.52 & 250 & 11.74 & ${}_{-0.18}^{+0.10}$ & ${}_{-0.38}^{+0.32}$ & 298.21 & ${}_{-0.22}^{+0.17}$ & ${}_{-0.57}^{+0.53}$ \\ 
		\hline 
		3 & $55355.49$\tablenotemark{b}  & 0.65 & 340 & 23.58 & ${}_{-1.18}^{+0.91}$ & ${}_{-4.13}^{+2.31}$ & 344.93 & ${}_{-1.04}^{+1.14}$ & ${}_{-2.90}^{+3.39}$ \\ 
		4 & $55370.74$\tablenotemark{b}  & 0.54 & 260 & 47.80 & ${}_{-0.22}^{+0.25}$ & ${}_{-0.48}^{+0.56}$ & 141.25 & ${}_{-0.16}^{+0.23}$ & ${}_{-0.45}^{+0.46}$ \\ 
		5 & $55387.54$\tablenotemark{b}  & 0.49 & 230 & 21.00 & ${}_{-0.59}^{+0.57}$ & ${}_{-1.56}^{+2.25}$ & 306.96 & ${}_{-1.12}^{+0.94}$ & ${}_{-2.28}^{+2.70}$ \\ 
		6 & $55421.51$\tablenotemark{b}  & 0.89 & 770 & 15.21 & ${}_{-3.10}^{+3.02}$ & ${}_{-7.41}^{+9.35}$ & 252.00 & ${}_{-6.48}^{+4.63}$ & ${}_{-16.65}^{+9.56}$ \\
		7 & $55464.90$\tablenotemark{b}  & 0.77 & 460 & 13.40 & ${}_{-0.15}^{+0.24}$ & ${}_{-0.45}^{+0.52}$ & 266.29 & ${}_{-0.23}^{+0.22}$ & ${}_{-0.62}^{+0.58}$ \\
		8 & $55478.38$\tablenotemark{b}  & 0.86 & 660 & 11.09 & ${}_{-0.19}^{+0.18}$ & ${}_{-0.49}^{+0.41}$ & 331.08 & ${}_{-0.35}^{+0.18}$ & ${}_{-0.80}^{+0.49}$ \\
		9 & $55497.30$\tablenotemark{b}  & 0.92 & 950 & 0.50 & ${}_{-0.10}^{+0.10}$ & ${}_{-0.21}^{+0.25}$  & 88.95  & ${}_{-0.25}^{+0.18}$ & ${}_{-0.53}^{+0.48}$ \\
		10 & $55513.60$\tablenotemark{b} & 0.80 & 520 & 3.15 & ${}_{-0.25}^{+0.33}$ & ${}_{-0.63}^{+0.70}$ & 285.95 & ${}_{-0.42}^{+0.58}$ & ${}_{-1.50}^{+1.29}$ \\ 
		11 & $55589.56$\tablenotemark{b} & 0.52 & 240 & 1.03 & ${}_{-0.08}^{+0.07}$ & ${}_{-0.21}^{+0.19}$ & 307.71 & ${}_{-0.08}^{+0.08}$ & ${}_{-0.44}^{+0.52}$ \\
		\hline
		12 & $55702.77$\tablenotemark{b} & 0.60 & 300 & 20.30 & ${}_{-0.62}^{+0.44}$ & ${}_{-1.43}^{+1.00}$ & 235.13 & ${}_{-0.55}^{+0.89}$ & ${}_{-1.76}^{+2.70}$ \\
		13 & $55722.43$\tablenotemark{b} & 0.47 & 210 & 35.55 & ${}_{-0.29}^{+0.28}$ & ${}_{-0.69}^{+0.69}$ & 272.22 & ${}_{-0.38}^{+0.50}$ & ${}_{-1.19}^{+1.23}$ \\ 
		14 & $55764.22$\tablenotemark{b} & 0.46 & 210 & 5.29 &  ${}_{-1.96}^{+1.87}$ & ${}_{-4.72}^{+4.85}$ & 315.66 & ${}_{-1.39}^{+2.37}$ & ${}_{-5.35}^{+5.91}$ \\ 
		15 & $55896.86$\tablenotemark{b} & 0.59 & 300 & 1.87 &  ${}_{-0.37}^{+0.57}$ & ${}_{-1.18}^{+1.25}$ & 222.87 & ${}_{-1.14}^{+0.90}$ & ${}_{-7.73}^{+1.95}$ \\ 
		16 & $55911.28$\tablenotemark{b} & 0.86 & 660 & 19.10 & ${}_{-0.77}^{+0.54}$ & ${}_{-2.21}^{+2.21}$ & 36.65 &  ${}_{-0.56}^{+0.61}$ & ${}_{-1.71}^{+1.85}$ \\
		\hline
		17 & 56062.96 & 0.45 & 200 & 31.96 & ${}_{-0.37}^{+0.30}$ & ${}_{-0.85}^{+0.81}$ & 198.74 & ${}_{-0.18}^{+0.49}$ & ${}_{-1.09}^{+1.44}$ \\
		18 & 56146.21 & 0.55 & 260 & 1.57 &  ${}_{-0.18}^{+0.22}$ & ${}_{-0.42}^{+0.46}$ & 330.10 & ${}_{-0.36}^{+0.24}$ & ${}_{-0.82}^{+0.65}$ \\
		19 & 56211.77 & 0.46 & 210 & -2.39 & ${}_{-0.19}^{+0.18}$ & ${}_{-0.51}^{+0.42}$ & 205.11 & ${}_{-0.24}^{+0.17}$ & ${}_{-0.66}^{+0.54}$ \\
		20 & 56226.60 & 0.88 & 750 & 28.04 & ${}_{-0.23}^{+0.31}$ & ${}_{-0.66}^{+0.67}$ & 169.61 & ${}_{-0.48}^{+0.45}$ & ${}_{-1.11}^{+1.16}$ \\
		21 & 56470.11\tablenotemark{c} & 0.87 & 670 & 14.46 & ${}_{-0.39}^{+0.40}$ & ${}_{-0.94}^{+0.86}$ & 93.38 & ${}_{-0.34}^{+0.33}$ & ${}_{-0.90}^{+0.83}$ \\
		22 & 56521.83 & 0.71 & 400 & -4.44 & ${}_{-0.39}^{+0.42}$ & ${}_{-0.94}^{+1.21}$ & 224.89 & ${}_{-0.32}^{+0.33}$ & ${}_{-1.19}^{+0.87}$ \\
		23 & 56579.91 & 0.49 & 390 & 10.20 & ${}_{-0.15}^{+0.15}$ & ${}_{-0.49}^{+0.34}$ & 32.94  & ${}_{-0.27}^{+0.20}$ & ${}_{-0.62}^{+0.63}$ \\
		24 & 56666.50 & 0.90 & 850 & 32.82 & ${}_{-0.14}^{+0.16}$ & ${}_{-0.41}^{+0.39}$ & 293.29 & ${}_{-0.40}^{+0.18}$ & ${}_{-1.08}^{+0.55}$ \\
		25 & 56799.96 & 0.73 & 400 & 18.05 & ${}_{-0.63}^{+0.75}$ & ${}_{-1.80}^{+1.94}$ & 349.39 & ${}_{-1.75}^{+1.13}$ & ${}_{-4.12}^{+2.89}$ \\
		26 & 56817.64 & 0.66 & 340 & 1.29  & ${}_{-0.29}^{+0.33}$ & ${}_{-0.74}^{+0.83}$ & 106.26 & ${}_{-0.74}^{+0.86}$ & ${}_{-1.90}^{+2.27}$ \\
		27 & 56819.20 & 0.995 & 4450 & 11.42 & ${}_{-0.08}^{+0.07}$ & ${}_{-0.17}^{+0.17}$ & 110.63 & ${}_{-0.28}^{+0.16}$ & ${}_{-0.55}^{+0.46}$ \\
		28 & 57049.48 & 0.46 & 210 & 4.56 &  ${}_{-0.12}^{+0.19}$ & ${}_{-0.50}^{+0.68}$ & 100.48 & ${}_{-0.34}^{+0.23}$ & ${}_{-1.87}^{+0.95}$ \\
		29 & 57157.94 & 0.52 & 240 & 12.18 & ${}_{-0.18}^{+0.19}$ & ${}_{-0.35}^{+0.37}$ & 91.60 & ${}_{-0.37}^{+0.10}$ & ${}_{-0.74}^{+0.16}$ \\
	\enddata
	\tablenotetext{a}{These events were included in \cite{IceCube:IC59NuMuDiffuse}.}
	\tablenotetext{b}{These events were included in \cite{IceCube:IC79NuMuDiffuse}.}
	\tablenotetext{c}{This event is identical to Event 38 in \cite{Aartsen:2015zva:hese}.}
\end{deluxetable*}

\subsection{Test for anisotropies related to the galactic plane}\label{sec:aniso:galactic:plane}

As discussed in Sec. \ref{sec:flux:discuss} the measurement in this paper
 confirms the observation of an all-sky diffuse high-energy 
astrophysical neutrino flux. However, 
a tension exists between the measured spectral index of this analysis with the starting event data
which originates mostly from the Southern hemisphere.
Furthermore, \cite{Neronov:2015osa} claim 
inconsistency of the previously published starting event data
with an isotropic signal with a preference of a galactic latitude dependency.
As the comparison to the Southern hemisphere is subject to different energy thresholds and detector systematics, we perform a simple,
self-consistent test for a dominant signal from the galactic plane.

We split the sample in two right ascension regions, one containing main parts of the galactic plane: $\alpha \in [0.0^\circ,108.9^\circ) \cup [275.0^\circ,360.0^\circ)$ and one excluding it: $\alpha \in [108.9^\circ,275.0^\circ)$.
These intervals are chosen such, that the two split samples are of similar statistics, resulting in 162363 and 189931 events respectively.
Both samples are fitted independently and the aforementioned systematics can be considered identical as they are equalized by the daily Earth rotation.

\begin{figure}
	\centering
	\includegraphics[width=1.\columnwidth]{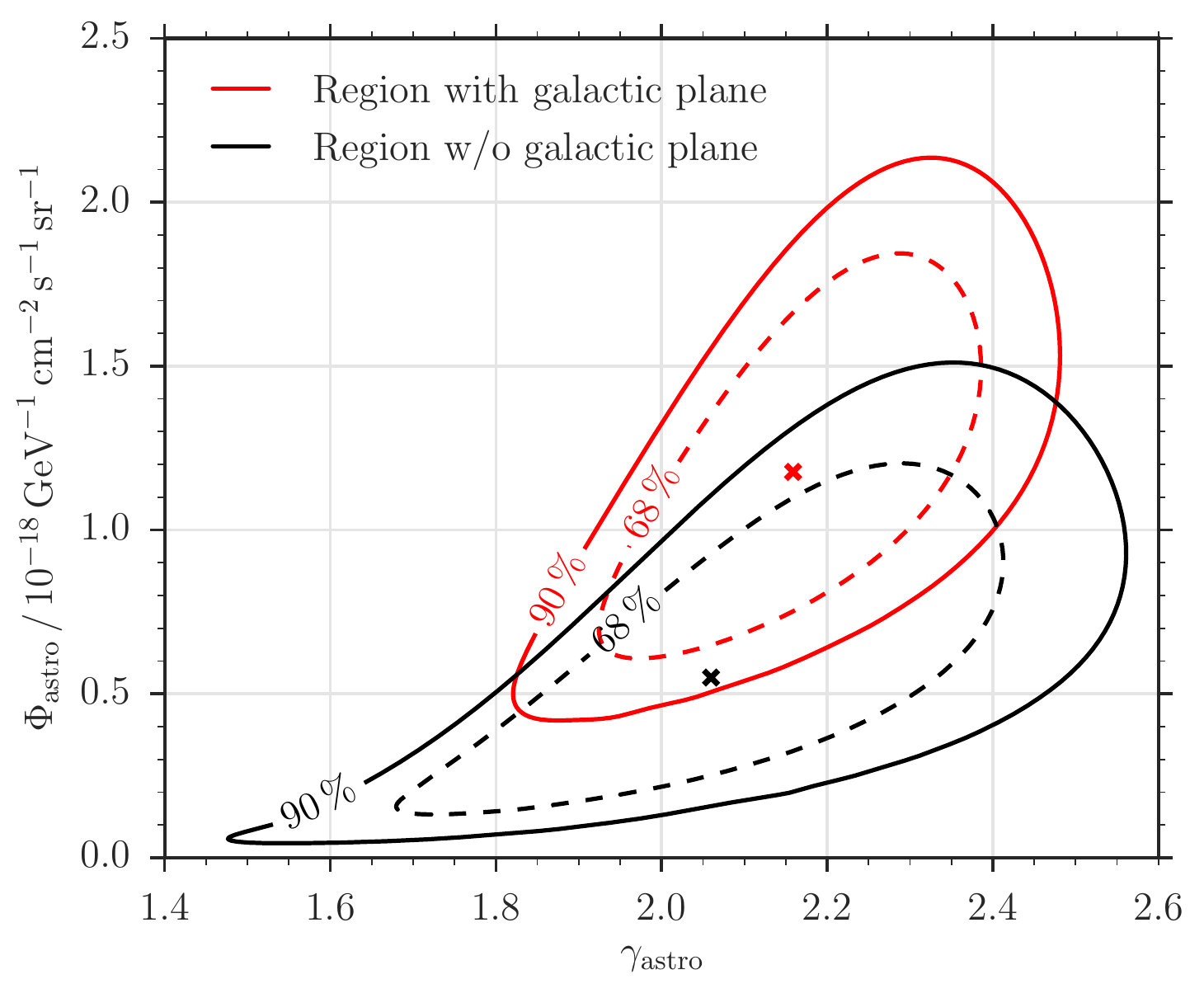}
	\caption{Two-dimensional profile likelihood scans of the astrophysical parameters $\Phi_\mathrm{astro}$ and $\gamma_\mathrm{astro}$ for the two disjoint right ascension regions, one containing the Northern Hemisphere part of the galactic plane (red) and the other not (black). The contour lines at $68\%$ and $90\%$ CL assume Wilks’ theorem. \label{fig:galactic_plane}}
\end{figure}

The fit results, shown in Fig.\ref{fig:galactic_plane}, 
is a small but not statistically significant larger flux and softer 
spectrum from the region including the galactic plane. 
The p-value for both results being compatible is at about $43\%$.
In conclusion, the observed flux is not dominated by the galactic plane.
However a small, sub-dominant contribution cannot be excluded.

\section{Search for a Signature of Prompt Atmospheric Neutrinos}\label{sec:prompt}
The expected prompt neutrino flux provides a background for the measurement of the astrophysical flux. However, a flux of prompt neutrinos is interesting by itself and can be constrained by the present analysis.
\begin{figure}
	\centering
	\includegraphics[width=1.\columnwidth]{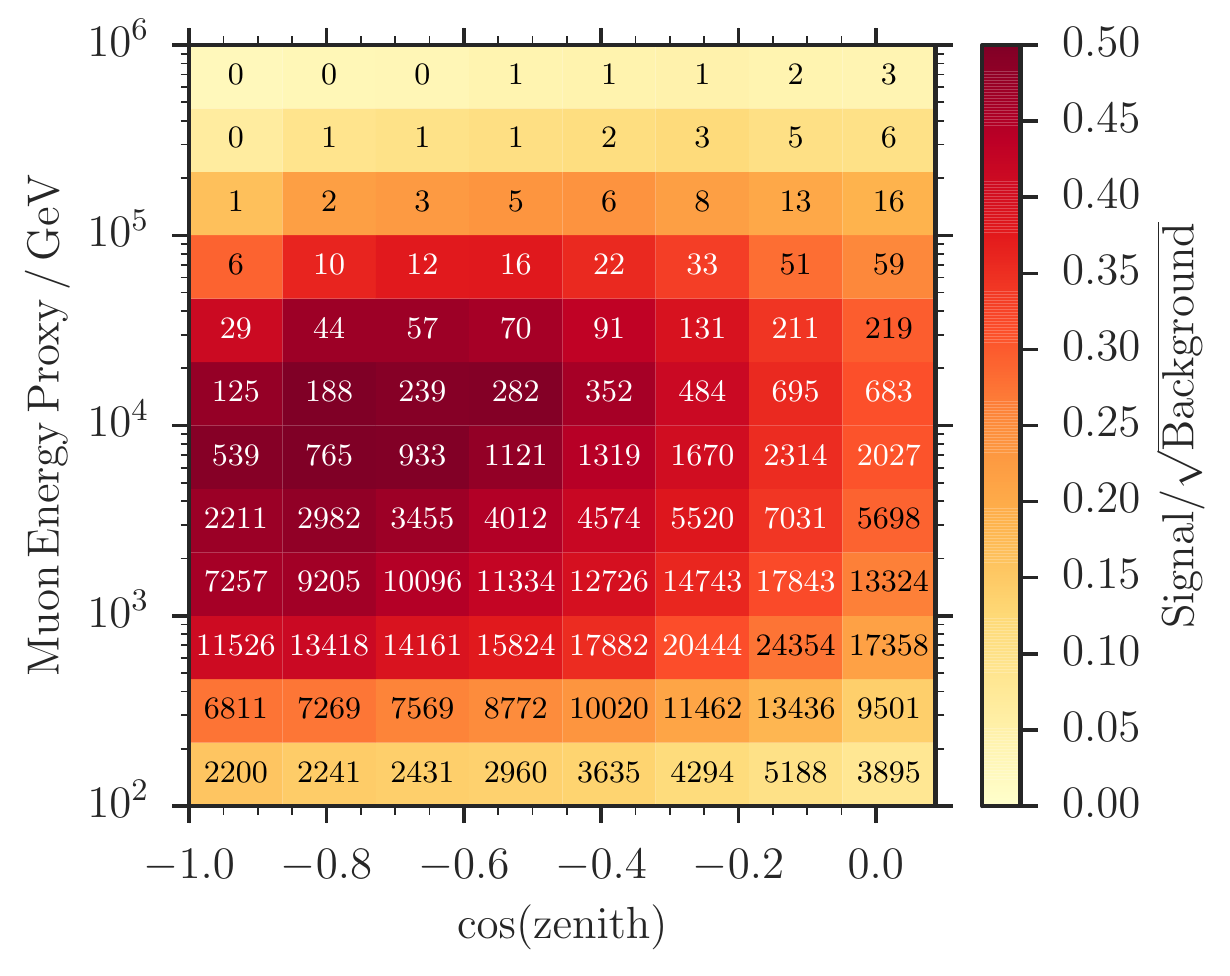}
	\caption{Signal over square root of background for the reconstructed muon energy vs. zenith angle corresponding to 6 years of IceCube data after applying the event selection for the 86-string configuration (IC2012-2014). Here, background is defined as the sum of the conventional atmospheric \citep{Conv:Honda2007} and astrophysical ($10^{-8}\times E^{-2}$) $\nu_\mu + \bar{\nu}_\mu$ flux. The prompt atmospheric \citep{Prompt:ERS} $\nu_\mu + \bar{\nu}_\mu$ flux is defined as signal. The numbers in each bin correspond to the expected number of background events in 6 years. \label{fig:2d_astro_signature}}
\end{figure}

The prompt flux predicted in \cite{Prompt:ERS} is sub-dominant to the conventional flux at low energies and the astrophysical flux at high energies. Nevertheless, the correlation of the energy spectrum and arrival directions of neutrinos at the detector lead to a clear signature.
Figure \ref{fig:2d_astro_signature} shows the pulls for simulated data corresponding to six years of live time and based on the IC2012-2014 event selection. Here, signal is defined as the prompt expectation and background is the sum of the conventional and astrophysical flux. The main effect of a prompt neutrino flux on the two observables will be visible for muon energy proxy values between $1\,$TeV and $100\,$TeV in the fairly up-going directions. However, a large part of this signature is absorbed within the uncertainties represented by the implemented nuisance parameters (cf. Sec. \ref{sec:systematics}).

\begin{figure}
	\centering
	\includegraphics[width=0.45\textwidth]{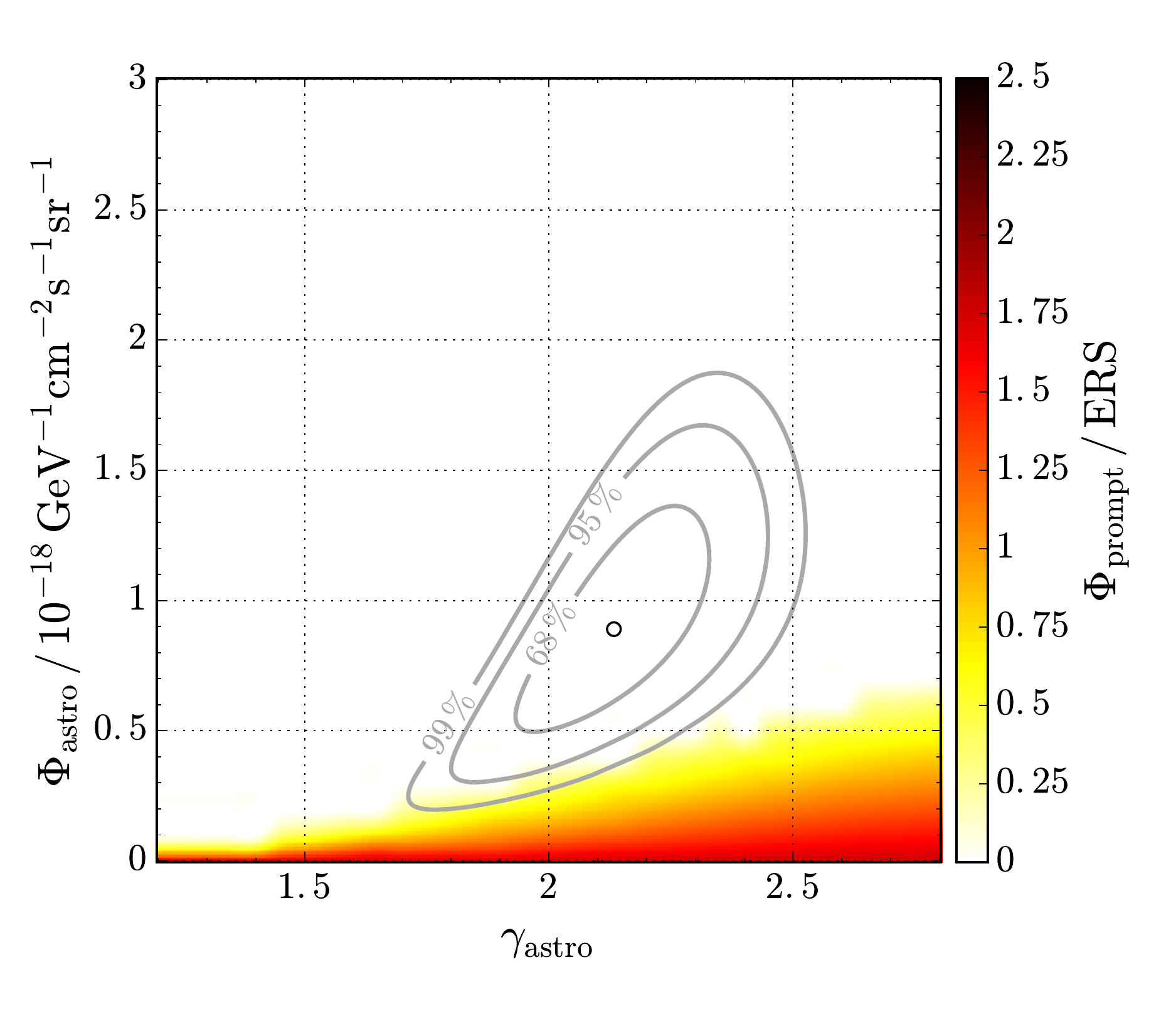}
	\caption{Best-fit prompt normalization $\Phi_\mathrm{prompt}$ in units of the model in \cite{Prompt:ERS} for each scan point $\Phi_\mathrm{astro}$, $\gamma_\mathrm{astro}$. Additionally, the two-dimensional contours for $\Phi_\mathrm{astro}$, $\gamma_\mathrm{astro}$ are shown. \label{fig:2dscan_spectator_prompt}}
\end{figure}

\begin{figure}
	\centering
	\includegraphics[width=1.\columnwidth]{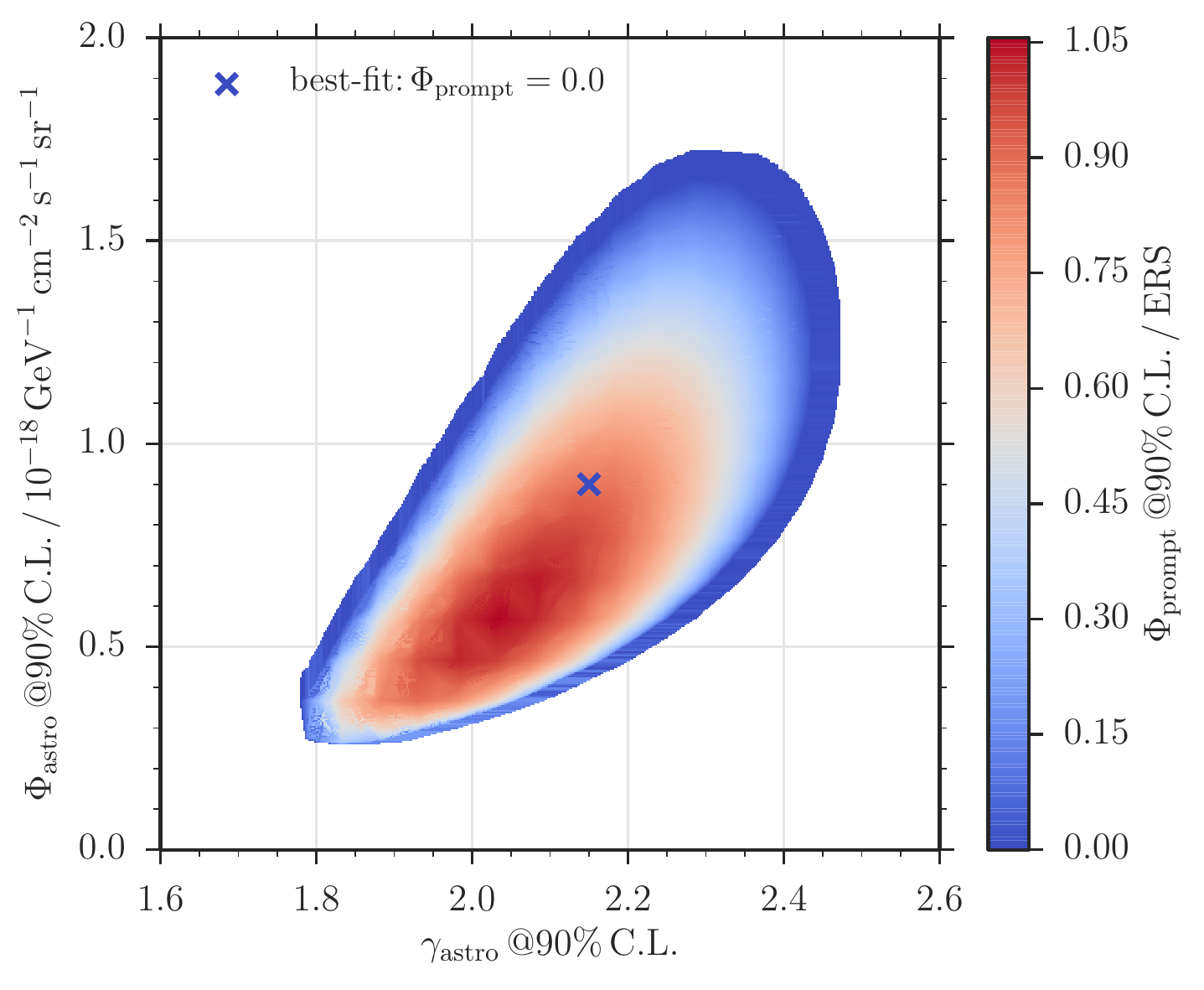}
	\caption{$90\%$ CL contour assuming Wilks' theorem based on a three dimensional profile likelihood scans of the astrophysical parameters $\Phi_\mathrm{astro}$, $\gamma_\mathrm{astro}$ and the prompt normalization $\Phi_\mathrm{prompt}$ in units of the model in \cite{Prompt:ERS}. \label{fig:3dscan}}
\end{figure}

The overall best-fit prompt normalization is zero. Figure \ref{fig:2dscan_spectator_prompt} shows the best-fit prompt normalization as a function of the astrophysical normalization and spectral index. Additionally, the two-dimensional confidence contours for the astrophysical parameters are shown. In the region where our experimental data is compatible with our single power law model,
 the best-fit prompt normalization does not deviate from zero. Only for strong deviations from the best-fit astrophysical spectrum is a non-zero prompt normalization fitted, but this is strongly disfavored with respect to the best-fit. Such behavior is expected. If the astrophysical flux decreases, the measured high-energy events need to be explained by another component.
Assuming an unbroken power law model for the astrophysical flux, the sensitivity for the prompt neutrino flux, taking into account the systematic uncertainties, is estimated to be $1.5\times\mathrm{ERS}$.
Note that the sensitivity (median expected upper limit in the absence of a prompt neutrino flux) on a prompt neutrino flux depends on the chosen input values for the astrophysical flux.

In the absence of an indication of a non-zero prompt contribution 
an upper limit is calculated.
Based on the profile likelihood for the prompt normalization, the upper limit at $90\%$ confidence level is $0.50\times\mathrm{ERS}$. The more stringent limit compared to the sensitivity is caused by an under-fluctuation of the conventional atmospheric and astrophysical background by about one standard deviation.

For this reason we scan the resulting limit on the prompt flux as a function
of the astrophysical signal parameters.

Figure \ref{fig:3dscan} shows the joint three-dimensional $90\%$ confidence region for the prompt flux and the astrophysical parameters. It was obtained using Wilks' theorem, and is bound by the surface for which $-2\Delta\log L$ is 6.25 higher than the best-fit value. The maximum prompt flux in the three-dimensional confidence region is $1.06\times\mathrm{ERS}$. We take this as a conservative upper limit on the prompt flux. Further tests have shown that reasonable changes to the astrophysical hypothesis, such as the introduction of a high-energy cut-off, have only small effects on this limit.

\begin{figure}
	\centering
	\includegraphics[width=1.\columnwidth]{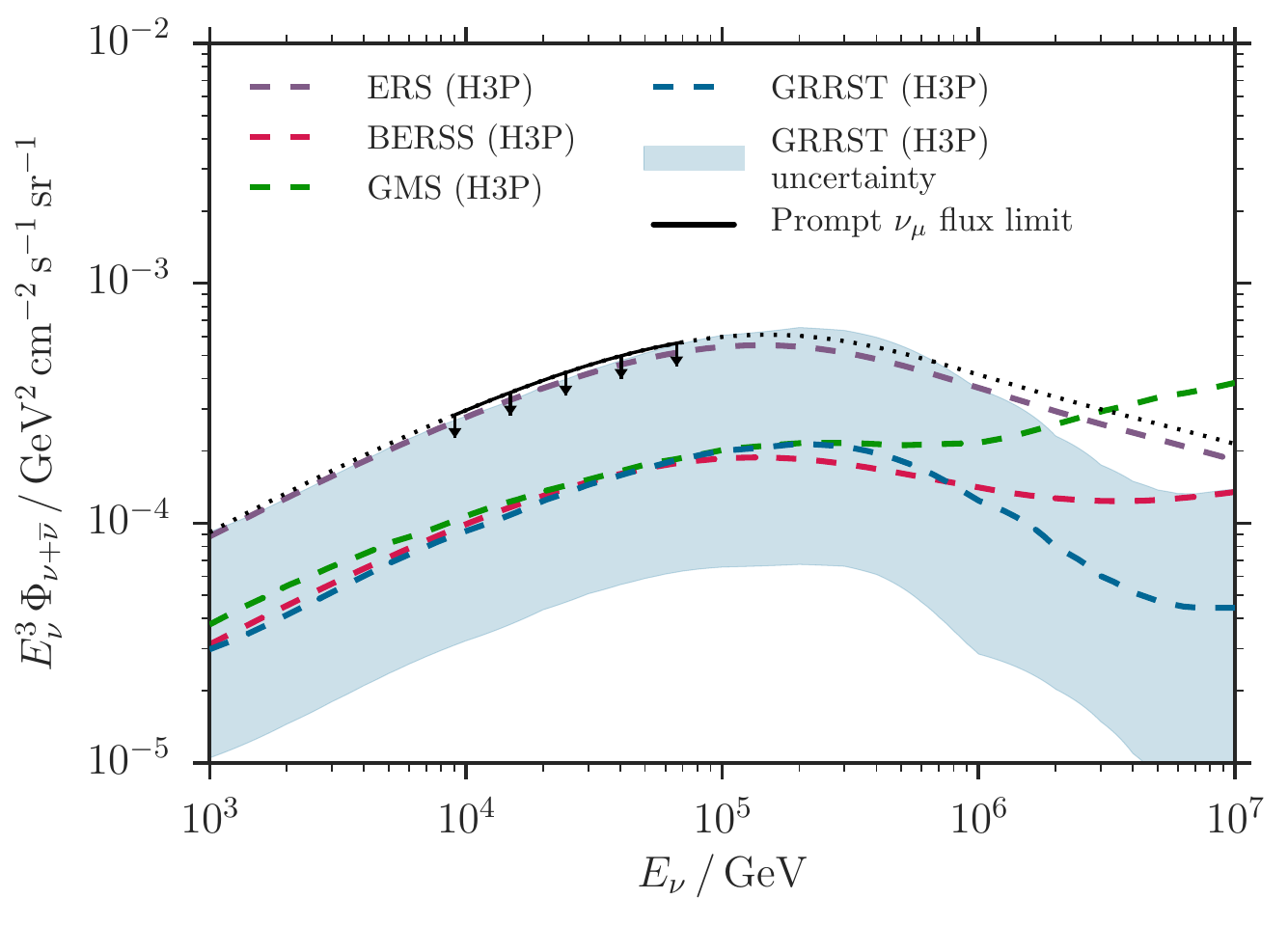}
	\caption{Prompt atmospheric muon neutrino flux predictions shown as dashed lines \citep{Prompt:ERS,Prompt:BERSS,Prompt:GRSST,Prompt:GMS} in comparison to the constraint on the prompt flux given by this analysis. The shaded area shows the uncertainty band corresponding to the prediction in \cite{Prompt:GRSST}. Besides the ERS (H3p) prediction this is the closest band to the prompt flux constraint. For a better readability the uncertainty bands of the other models are not shown. The black solid line shows the neutrino energy region where the prompt neutrino flux based on the model in \cite{Prompt:ERS} is constrained. The black dotted line indicates the model behavior including the best-fit nuisance parameters beyond the sensitive energy range. All flux predictions are based on the cosmic ray model from \cite{Gaisser:2012zz}. \label{fig:flux_limit_comparisons}}
\end{figure}

Several more recent calculations of the prompt flux have been published: GMS (H3p) \citep{Prompt:GMS}, BERSS (H3p) \citep{Prompt:BERSS} and GRRST (H3p) \citep{Prompt:GRSST}. Figure \ref{fig:flux_limit_comparisons} shows multiple predictions for the prompt flux as well as the upper limit calculated here using the prediction from \cite{Prompt:ERS} and taking into account a more realistic cosmic-ray model \citep{Gaisser:2012zz}. Since nuisance parameters describing the uncertainties of the cosmic-ray model, e.g. the cosmic-ray spectral index, are implemented the upper limit curve slightly deviates from the ERS prediction including the knee.
The energy range has been calculated such that the limit increases by 10\% if only neutrinos with energies in that range are taken into account. For the sensitive region which is between $9\,\mathrm{TeV}$ to $69\,\mathrm{TeV}$ the effect of the prompt predictions is only a change in normalization and it is therefore appropriate to convert the limit obtained with the ERS prediction to the other predictions. Also the cosmic ray composition only changes the normalization in this energy range. The values are summarized in Tab. \ref{tab:prompt:limits}.

\begin{deluxetable}{p{0.4\columnwidth}p{0.2\columnwidth}}
	\tablewidth{1.0\columnwidth}
	\tablecaption{Limits for fluxes of prompt neutrinos for different predictions. The limits for GMS (H3p) \citep{Prompt:GMS}, BERSS (H3p) \citep{Prompt:BERSS} and GRRST (H3p) \citep{Prompt:GRSST} are determined by rescaling the ERS (H3p) limit with the corresponding flux ratio at $30\,\mathrm{TeV}$ which is well within the sensitive energy range. All flux predictions are based on the cosmic ray model from \cite{Gaisser:2012zz}. \label{tab:prompt:limits}}
	\tablehead{\colhead{Model} & \colhead{Flux limit}}
	\startdata
		ERS (H3p) &  1.06 \\
		GMS (H3p) & $\approx 2.9$ \\
		BERSS (H3p) & $\approx 3.0$ \\
		GRRST (H3p) & $\approx 3.1$ \\
	\enddata
\end{deluxetable}

\newpage

\section{Summary and Conclusions}

In this paper we have presented the result of analyzing 6 years of up-going muon data measured with the IceCube neutrino telescope. We measure an astrophysical
flux of $	\Phi_{\nu+\overline{\nu}} = \left(0.90^{+0.30}_{-0.27}\right) 10^{-18}\,\mathrm{GeV^{-1}\,cm^{-2}\,sr^{-1}\,s^{-1}} \cdot (E_\nu/100\,\mathrm{TeV})^{-(2.13 \pm 0.13)} $ with statistical significance of $5.6$ standard deviations with respect to only being of atmospheric origin.
With this result we have further established the observation of an astrophysical neutrino 
signal \citep{Aartsen:2013jdh,IceCube:HESE3Years, IceCube:IC79NuMuDiffuse}
in a second, largely independent detection channel. 
The detection channel used here is of great interest because of the good directional reconstruction of detected muons and a large signal efficiency with an estimated number of about $500$ astrophysical neutrinos included in this data sample.

The data include an exceptionally high-energy muon with $(2.6 \pm 0.3)$\,PeV deposited energy, which is the highest energy lepton that has been reported to date.

A parametric unfolding of neutrino energies shows that the spectrum extends to about $10$\,PeV in neutrino energy with no significant spectral break or cut-off.

The measured hard spectral index of $\gamma = 2.13 \pm 0.13$ is in tension with 
complementary measurements of IceCube, which have a lower energy threshold by about one order of magnitude and are predominantly sensitive to the Southern hemisphere. However, the 
consistency of the observed fluxes at high energies may be interpreted 
as indication of a spectral break or additional astrophysical component at 
lower energy to which this analysis is not sensitive.

For the highest-energy events no correlation with known high-energy gamma-ray 
sources or other astrophysical objects could be identified.

By splitting the data in right ascension, we 
find no significant correlation with the orientation of the galactic plane and
conclude that the dominant fraction of the flux is largely all-sky and isotropic.

The present analysis is also 
sensitive to a flux of prompt neutrinos which are expected from the decay of heavy mesons in the atmosphere. We find no indications for such a signal. 
However, because the prompt flux is subdominant to the astrophysical and conventional atmospheric 
neutrino flux, the exclusion depends on the assumed astrophysical model parameters. Variations
of the astrophysical flux uncertainties lead to a conservative 
exclusion limit of approximately at the level of the mean expected 
flux normalization from \cite{Prompt:ERS}.
 For the first time, it is possible to constrain such a flux 
in this range of theoretical predictions.
However, recent perturbative QCD calculations from \cite{Prompt:GMS}, \cite{Prompt:BERSS} and \citep{Prompt:GRSST} predict lower prompt neutrino fluxes which are not yet constrained by the upper limit.

\acknowledgments

We acknowledge the support from the following agencies:
U.S. National Science Foundation-Office of Polar Programs,
U.S. National Science Foundation-Physics Division,
University of Wisconsin Alumni Research Foundation,
the Grid Laboratory Of Wisconsin (GLOW) grid infrastructure at the University of Wisconsin - Madison, the Open Science Grid (OSG) grid infrastructure;
U.S. Department of Energy, and National Energy Research Scientific Computing Center,
the Louisiana Optical Network Initiative (LONI) grid computing resources;
Natural Sciences and Engineering Research Council of Canada,
WestGrid and Compute/Calcul Canada;
Swedish Research Council,
Swedish Polar Research Secretariat,
Swedish National Infrastructure for Computing (SNIC),
and Knut and Alice Wallenberg Foundation, Sweden;
German Ministry for Education and Research (BMBF),
Deutsche Forschungsgemeinschaft (DFG),
Helmholtz Alliance for Astroparticle Physics (HAP),
Research Department of Plasmas with Complex Interactions (Bochum), Germany;
Fund for Scientific Research (FNRS-FWO),
FWO Odysseus programme,
Flanders Institute to encourage scientific and technological research in industry (IWT),
Belgian Federal Science Policy Office (Belspo);
University of Oxford, United Kingdom;
Marsden Fund, New Zealand;
Australian Research Council;
Japan Society for Promotion of Science (JSPS);
the Swiss National Science Foundation (SNSF), Switzerland;
National Research Foundation of Korea (NRF);
Villum Fonden, Danish National Research Foundation (DNRF), Denmark


\begin{thebibliography}{}
\providecommand\natexlab[1]{#1}
\providecommand\JournalTitle[1]{#1}

\bibitem[{Aartsen {et~al.}(2015{\natexlab{a}})Aartsen, Ackermann, Adams,
  Aguilar, Ahlers, {et~al.}}]{Aartsen:2016GRB1}
Aartsen, M.~G., Ackermann, M., Adams, J., {et~al.} 2015{\natexlab{a}},
  \href{http://stacks.iop.org/2041-8205/805/i=1/a=L5}{\JournalTitle{The
  Astrophysical Journal Letters}, 805, L5}

\bibitem[{Aartsen {et~al.}(2013{\natexlab{a}})}]{Aartsen:2013jdh}
Aartsen, M.~G., {et~al.} 2013{\natexlab{a}},
  \href{http://dx.doi.org/10.1126/science.1242856}{\JournalTitle{Science}, 342,
  1242856}

\bibitem[{Aartsen
  {et~al.}(2013{\natexlab{b}})}]{IceCube:OpticalIcePropertiesPapers}
---. 2013{\natexlab{b}},
  \href{http://dx.doi.org/10.1016/j.nima.2013.01.054}{\JournalTitle{Nucl.Instrum.Meth.},
  A711, 73}

\bibitem[{Aartsen {et~al.}(2014{\natexlab{a}})}]{Aartsen:2013vja}
---. 2014{\natexlab{a}},
  \href{http://dx.doi.org/10.1088/1748-0221/9/03/P03009}{\JournalTitle{JINST},
  9, P03009}

\bibitem[{Aartsen {et~al.}(2014{\natexlab{b}})}]{IceCube:HESE3Years}
---. 2014{\natexlab{b}},
  \href{http://dx.doi.org/10.1103/PhysRevLett.113.101101}{\JournalTitle{Phys.Rev.Lett.},
  113, 101101}

\bibitem[{Aartsen {et~al.}(2014{\natexlab{c}})}]{IceCube:IC59NuMuDiffuse}
---. 2014{\natexlab{c}},
  \href{http://dx.doi.org/10.1103/PhysRevD.89.062007}{\JournalTitle{Phys.Rev.},
  D89, 062007}

\bibitem[{Aartsen {et~al.}(2015{\natexlab{b}})}]{Aartsen:2015knd}
---. 2015{\natexlab{b}},
  \href{http://dx.doi.org/10.1088/0004-637X/809/1/98}{\JournalTitle{Astrophys.
  J.}, 809, 98}

\bibitem[{Aartsen {et~al.}(2015{\natexlab{c}})}]{IceCube:MESE2Years}
---. 2015{\natexlab{c}},
  \href{http://dx.doi.org/10.1103/PhysRevD.91.022001}{\JournalTitle{Phys. Rev.
  D}, 91, 022001}

\bibitem[{Aartsen {et~al.}(2015{\natexlab{d}})}]{IceCube:IC79NuMuDiffuse}
---. 2015{\natexlab{d}},
  \href{http://dx.doi.org/10.1103/PhysRevLett.115.081102}{\JournalTitle{Phys.
  Rev. Lett.}, 115, 081102}

\bibitem[{Aartsen {et~al.}(2015{\natexlab{e}})}]{IceCube:TD:PS:2008-2012}
---. 2015{\natexlab{e}},
  \href{http://dx.doi.org/10.1088/0004-637X/807/1/46}{\JournalTitle{Astrophys.
  J.}, 807, 46}

\bibitem[{Aartsen {et~al.}(2016{\natexlab{a}})}]{IceCube:PS7yr}
---. 2016{\natexlab{a}}, in preparation for Astrophys. J.

\bibitem[{Aartsen {et~al.}(2016{\natexlab{b}})}]{Aartsen:2016GRB2}
---. 2016{\natexlab{b}}, \href{http://arxiv.org/abs/1601.06484}{{\sffamily
  arXiv:1601.06484 [astro-ph.HE]}}

\bibitem[{Abbasi {et~al.}(2009)}]{Abbasi:2008aa}
Abbasi, R., {et~al.} 2009,
  \href{http://dx.doi.org/10.1016/j.nima.2009.01.001}{\JournalTitle{Nucl.
  Instrum. Meth.}, A601, 294}

\bibitem[{Abbasi {et~al.}(2010)}]{Abbasi:2010vc}
---. 2010,
  \href{http://dx.doi.org/10.1016/j.nima.2010.03.102}{\JournalTitle{Nucl.
  Instrum. Meth.}, A618, 139}

\bibitem[{Abbasi {et~al.}(2013)}]{EnergyReco:truncated}
---. 2013,
  \href{http://dx.doi.org/10.1016/j.nima.2012.11.081}{\JournalTitle{Nucl.
  Instrum. Meth. A}, 703, 190}

\bibitem[{Acero {et~al.}(2015)}]{Fermi3FGL}
Acero, F., {et~al.} 2015, \href{http://arxiv.org/abs/1501.02003}{{\sffamily
  arXiv:1501.02003 [astro-ph.HE]}}

\bibitem[{Achterberg {et~al.}(2006)}]{IceCube:FirstYear}
Achterberg, A., {et~al.} 2006,
  \href{http://dx.doi.org/10.1016/j.astropartphys.2006.06.007}{\JournalTitle{Astropart.Phys.},
  26, 155}

\bibitem[{Ackermann {et~al.}(2006)Ackermann, Ahrens, Bai, Bartelt, Barwick,
  {et~al.}}]{ackermann2006optical}
Ackermann, M., Ahrens, J., Bai, X., {et~al.} 2006,
  \href{http://dx.doi.org/10.1029/2005JD006687}{\JournalTitle{Journal of
  Geophysical Research: Atmospheres}, 111}, d13203

\bibitem[{Ageron {et~al.}(2011)}]{antares:2011nsa}
Ageron, M., {et~al.} 2011,
  \href{http://dx.doi.org/10.1016/j.nima.2011.06.103}{\JournalTitle{Nucl.Instrum.Meth.},
  A656, 11}

\bibitem[{Ahrens {et~al.}(2004)}]{Ahrens:2003fg}
Ahrens, J., {et~al.} 2004,
  \href{http://dx.doi.org/10.1016/j.nima.2004.01.065}{\JournalTitle{Nucl.Instrum.Meth.},
  A524, 169}

\bibitem[{{AIRS Science Team/Joao Texeira}(2013)}]{AQUA}
{AIRS Science Team/Joao Texeira}. 2013, {Aqua AIRS Level 3 Daily Standard
  Physical Retrieval (AIRS+AMSU),version 006}, Tech. rep., Greenbelt, MD, USA,
  NASA Goddard Earth Science Data and Information Services Center (GES DISC)

\bibitem[{Anchordoqui {et~al.}(2014)Anchordoqui, Goldberg, Lynch, Olinto, Paul,
  \& Weiler}]{PhysRevD.89.083003}
Anchordoqui, L.~A., Goldberg, H., Lynch, M.~H., {et~al.} 2014,
  \href{http://dx.doi.org/10.1103/PhysRevD.89.083003}{\JournalTitle{Phys. Rev.
  D}, 89, 083003}

\bibitem[{Andres {et~al.}(2000)Andres, Askebjer, Barwick, Bay, Bergstrom,
  {et~al.}}]{Andres:1999hm}
Andres, E., Askebjer, P., Barwick, S., {et~al.} 2000,
  \href{http://dx.doi.org/10.1016/S0927-6505(99)00092-4}{\JournalTitle{Astropart.Phys.},
  13, 1}

\bibitem[{Athar {et~al.}(2006)Athar, Kim, \& Lee}]{Athar:Oscillation}
Athar, H., Kim, C.~S., \& Lee, J. 2006,
  \href{http://dx.doi.org/10.1142/S021773230602038X}{\JournalTitle{Mod. Phys.
  Lett.}, A21, 1049}

\bibitem[{Bechtol {et~al.}(2015)Bechtol, Ahlers, Di~Mauro, Ajello, \&
  Vandenbroucke}]{Bechtol:2015}
Bechtol, K., Ahlers, M., Di~Mauro, M., Ajello, M., \& Vandenbroucke, J. 2015,
  \href{http://arxiv.org/abs/1511.00688}{{\sffamily arXiv:1511.00688
  [astro-ph.HE]}}

\bibitem[{Becker(2008)}]{Becker:2007sv}
Becker, J.~K. 2008,
  \href{http://dx.doi.org/10.1016/j.physrep.2007.10.006}{\JournalTitle{Phys.
  Rept.}, 458, 173}

\bibitem[{Bell(2013)}]{Bell201356}
Bell, A. 2013,
  \href{http://dx.doi.org/http://dx.doi.org/10.1016/j.astropartphys.2012.05.022}{\JournalTitle{Astroparticle
  Physics}, 43, 56 }

\bibitem[{Belolaptikov {et~al.}(1997)}]{Belolaptikov:1997ry}
Belolaptikov, I., {et~al.} 1997,
  \href{http://dx.doi.org/10.1016/S0927-6505(97)00022-4}{\JournalTitle{Astropart.Phys.},
  7, 263}

\bibitem[{Bhattacharya {et~al.}(2015)Bhattacharya, Enberg, Reno, Sarcevic, \&
  Stasto}]{Prompt:BERSS}
Bhattacharya, A., Enberg, R., Reno, M.~H., Sarcevic, I., \& Stasto, A. 2015,
  \href{http://dx.doi.org/10.1007/JHEP06(2015)110}{\JournalTitle{Journal of
  High Energy Physics}, 2015, 1}

\bibitem[{Boerner {et~al.}(2015)Boerner, Ruhe, Scheriau, \&
  Schmitz}]{Aartsen:2015zva:unfolding}
Boerner, M., Ruhe, T., Scheriau, F., \& Schmitz, M. 2015,
  \href{https://inspirehep.net/record/1398539/files/arXiv:1510.05223.pdf}{in
  {Proceedings, 34th International Cosmic Ray Conference (ICRC 2015)}}, 53

\bibitem[{Chirkin(2013{\natexlab{a}})}]{LLH:Dima}
Chirkin, D. 2013{\natexlab{a}}, \JournalTitle{ArXiv e-prints},
  \href{http://arxiv.org/abs/1304.0735}{{\sffamily arXiv:1304.0735
  [astro-ph.IM]}}

\bibitem[{Chirkin(2013{\natexlab{b}})}]{Chirkin:PPC}
---. 2013{\natexlab{b}},
  \href{http://dx.doi.org/10.1016/j.nima.2012.11.170}{\JournalTitle{Nucl.
  Instrum. Meth.}, A725, 141}

\bibitem[{Chirkin \& Rhode(2004)}]{MMC}
Chirkin, D., \& Rhode, W. 2004,
  \href{http://arxiv.org/abs/hep-ph/0407075}{{\sffamily arXiv:hep-ph/0407075
  [hep-ph]}}

\bibitem[{Chirkin {et~al.}(2014)}]{Aartsen:2013ola:chirkin}
Chirkin, D., {et~al.} 2014,
  \href{https://inspirehep.net/record/1255632/files/arXiv:1309.7010.pdf}{in
  {Proceedings, 33rd International Cosmic Ray Conference (ICRC2013): Rio de
  Janeiro, Brazil, July 2-9, 2013}}, 17

\bibitem[{Cholis \& Hooper(2013)}]{JCAP.2012.06.030}
Cholis, I., \& Hooper, D. 2013,
  \href{http://stacks.iop.org/1475-7516/2013/i=06/a=030}{\JournalTitle{Journal
  of Cosmology and Astroparticle Physics}, 2013, 030}

\bibitem[{Cooper-Sarkar {et~al.}(2011)Cooper-Sarkar, Mertsch, \& Sarkar}]{CSMS}
Cooper-Sarkar, A., Mertsch, P., \& Sarkar, S. 2011,
  \href{http://dx.doi.org/10.1007/JHEP08(2011)042}{\JournalTitle{JHEP}, 08,
  042}

\bibitem[{Desiati {et~al.}(2014)}]{ICRC:TemperatureVariation}
Desiati, P., {et~al.} 2014,
  \href{http://www.cbpf.br/%7Eicrc2013/papers/icrc2013-0492.pdf}{in
  {Proceedings, 33rd International Cosmic Ray Conference (ICRC2013): Rio de
  Janeiro, Brazil, July 2-9, 2013}}, 0492

\bibitem[{Enberg {et~al.}(2008)Enberg, Reno, \& Sarcevic}]{Prompt:ERS}
Enberg, R., Reno, M.~H., \& Sarcevic, I. 2008,
  \href{http://dx.doi.org/10.1103/PhysRevD.78.043005}{\JournalTitle{Phys.Rev.},
  D78, 043005}

\bibitem[{Freund \& Schapire(1997)}]{Freund:1997xna}
Freund, Y., \& Schapire, R.~E. 1997,
  \href{http://dx.doi.org/10.1006/jcss.1997.1504}{\JournalTitle{J. Comput.
  Syst. Sci.}, 55, 119}

\bibitem[{Gaisser(1990)}]{GaisserTextbook}
Gaisser, T.~K. 1990, {Cosmic Rays and Particle Physics} (Cambridge University
  Press)

\bibitem[{Gaisser(2012)}]{Gaisser:2012zz}
---. 2012,
  \href{http://dx.doi.org/10.1016/j.astropartphys.2012.02.010}{\JournalTitle{Astropart.Phys.},
  35, 801}

\bibitem[{Gaisser {et~al.}(1995)Gaisser, Halzen, \&
  Stanev}]{Gaisser:Halzen:Stanev}
Gaisser, T.~K., Halzen, F., \& Stanev, T. 1995,
  \href{http://dx.doi.org/10.1016/0370-1573(95)00003-Y}{\JournalTitle{Phys.Rept.},
  258, 173}

\bibitem[{Garzelli {et~al.}(2015)Garzelli, Moch, \& Sigl}]{Prompt:GMS}
Garzelli, M.~V., Moch, S., \& Sigl, G. 2015,
  \href{http://dx.doi.org/10.1007/JHEP10(2015)115}{\JournalTitle{Journal of
  High Energy Physics}, 2015, 1}

\bibitem[{Gauld {et~al.}(2016)Gauld, Rojo, Rottoli, Sarkar, \&
  Talbert}]{Prompt:GRSST}
Gauld, R., Rojo, J., Rottoli, L., Sarkar, S., \& Talbert, J. 2016,
  \href{http://dx.doi.org/10.1007/JHEP02(2016)130}{\JournalTitle{JHEP}, 02,
  130}

\bibitem[{Gazizov \& Kowalski(2005)}]{ANIS}
Gazizov, A., \& Kowalski, M.~P. 2005,
  \href{http://dx.doi.org/10.1016/j.cpc.2005.03.113}{\JournalTitle{Comput.
  Phys. Commun.}, 172, 203}

\bibitem[{{Gl\"usenkamp}(2015)}]{Glusenkamp:2015jca}
{Gl\"usenkamp}, T. 2015,
  \href{https://inspirehep.net/record/1343949/files/arXiv:1502.03104.pdf}{in
  {5th Roma International Conference on Astro-Particle physics (RICAP 14) Noto,
  Sicily, Italy, September 30-October 3, 2014}}

\bibitem[{Gonzalez-Garcia {et~al.}(2014)Gonzalez-Garcia, Halzen, \&
  Niro}]{GonzalezGarcia201439}
Gonzalez-Garcia, M., Halzen, F., \& Niro, V. 2014,
  \href{http://dx.doi.org/http://dx.doi.org/10.1016/j.astropartphys.2014.04.001}{\JournalTitle{Astroparticle
  Physics}, 57–58, 39 }

\bibitem[{Greisen(1960)}]{Greisen:1960}
Greisen, K. 1960,
  \href{http://dx.doi.org/10.1146/annurev.ns.10.120160.000431}{\JournalTitle{Ann.
  Rev. Nucl. Part. Sci.}, 10, 63}

\bibitem[{He {et~al.}(2013)He, Wang, Fan, Liu, \& Wei}]{PhysRevD.87.063011}
He, H.-N., Wang, T., Fan, Y.-Z., Liu, S.-M., \& Wei, D.-M. 2013,
  \href{http://dx.doi.org/10.1103/PhysRevD.87.063011}{\JournalTitle{Phys. Rev.
  D}, 87, 063011}

\bibitem[{Heck {et~al.}(1998)Heck, Schatz, Thouw, Knapp, \&
  Capdevielle}]{Heck:1998vt}
Heck, D., Schatz, G., Thouw, T., Knapp, J., \& Capdevielle, J. 1998, {CORSIKA:
  A Monte Carlo code to simulate extensive air showers}, Tech. rep.,
  Forschungszentrum Karlsruhe, Germany

\bibitem[{Hoerandel(2003)}]{hoerandel2003knee}
Hoerandel, J.~R. 2003, \JournalTitle{Astroparticle Physics}, 19, 193

\bibitem[{Honda {et~al.}(2007)Honda, Kajita, Kasahara, Midorikawa, \&
  Sanuki}]{Conv:Honda2007}
Honda, M., Kajita, T., Kasahara, K., Midorikawa, S., \& Sanuki, T. 2007,
  \href{http://dx.doi.org/10.1103/PhysRevD.75.043006}{\JournalTitle{Phys.Rev.},
  D75, 043006}

\bibitem[{Illana {et~al.}(2011)Illana, Lipari, Masip, \&
  Meloni}]{Illana:2010gh}
Illana, J.~I., Lipari, P., Masip, M., \& Meloni, D. 2011,
  \href{http://dx.doi.org/10.1016/j.astropartphys.2011.01.001}{\JournalTitle{Astropart.
  Phys.}, 34, 663}

\bibitem[{Kalashev {et~al.}(2013)Kalashev, Kusenko, \&
  Essey}]{PhysRevLett.111.041103}
Kalashev, O.~E., Kusenko, A., \& Essey, W. 2013,
  \href{http://dx.doi.org/10.1103/PhysRevLett.111.041103}{\JournalTitle{Phys.
  Rev. Lett.}, 111, 041103}

\bibitem[{Kashti \& Waxman(2005)}]{Kashti:2005qa}
Kashti, T., \& Waxman, E. 2005,
  \href{http://dx.doi.org/10.1103/PhysRevLett.95.181101}{\JournalTitle{Phys.
  Rev. Lett.}, 95, 181101}

\bibitem[{Klein {et~al.}(2013)Klein, Mikkelsen, \& Becker~Tjus}]{Klein:2012ug}
Klein, S.~R., Mikkelsen, R.~E., \& Becker~Tjus, J. 2013,
  \href{http://dx.doi.org/10.1088/0004-637X/779/2/106}{\JournalTitle{Astrophys.
  J.}, 779, 106}

\bibitem[{Koehne {et~al.}(2013)Koehne, Frantzen, Schmitz, Fuchs, Rhode,
  Chirkin, \& Becker~Tjus}]{PROPOSAL}
Koehne, J.~H., Frantzen, K., Schmitz, M., {et~al.} 2013,
  \href{http://dx.doi.org/10.1016/j.cpc.2013.04.001}{\JournalTitle{Comput.
  Phys. Commun.}, 184, 2070}

\bibitem[{Kopper(2011)}]{Kopper:CLSIM}
Kopper, C. 2011, \JournalTitle{GitHub: https://github.com/claudiok/clsim}

\bibitem[{Kopper {et~al.}(2015)Kopper, Kurahashi,
  {et~al.}}]{Aartsen:2015zva:hese}
Kopper, C., Kurahashi, N., {et~al.} 2015,
  \href{https://inspirehep.net/record/1398539/files/arXiv:1510.05223.pdf}{in
  {Proceedings, 34th International Cosmic Ray Conference (ICRC 2015)}}, 45

\bibitem[{Kotera {et~al.}(2010)Kotera, Allard, \& Olinto}]{Kotera:2010}
Kotera, K., Allard, D., \& Olinto, A.~V. 2010,
  \href{http://dx.doi.org/10.1088/1475-7516/2010/10/013}{\JournalTitle{JCAP},
  1010, 013}

\bibitem[{Laha {et~al.}(2013)Laha, Beacom, Dasgupta, Horiuchi, \&
  Murase}]{PhysRevD.88.043009}
Laha, R., Beacom, J.~F., Dasgupta, B., Horiuchi, S., \& Murase, K. 2013,
  \href{http://dx.doi.org/10.1103/PhysRevD.88.043009}{\JournalTitle{Phys. Rev.
  D}, 88, 043009}

\bibitem[{Lahav {et~al.}(2000)Lahav, Santiago, Webster, Strauss, Davis,
  Dressler, \& Huchra}]{ISI:000085634700018}
Lahav, O., Santiago, B., Webster, A., {et~al.} 2000,
  \href{http://dx.doi.org/10.1046/j.1365-8711.2000.03145.x}{\JournalTitle{MNRAS},
  312, 166}

\bibitem[{Lai {et~al.}(2000)Lai, Huston, Kuhlmann, Morfin, Olness, Owens,
  Pumplin, \& Tung}]{CTEQ5}
Lai, H.~L., Huston, J., Kuhlmann, S., {et~al.} 2000,
  \href{http://dx.doi.org/10.1007/s100529900196}{\JournalTitle{Eur. Phys. J.},
  C12, 375}

\bibitem[{Learned \& Mannheim(2000)}]{Learned:Mannheim}
Learned, J., \& Mannheim, K. 2000,
  \href{http://dx.doi.org/10.1146/annurev.nucl.50.1.679}{\JournalTitle{Ann.
  Rev. Nucl. Part. Sci.}, 50, 679}

\bibitem[{Learned \& Pakvasa(1995)}]{Learned:Oscillation}
Learned, J.~G., \& Pakvasa, S. 1995,
  \href{http://dx.doi.org/http://dx.doi.org/10.1016/0927-6505(94)00043-3}{\JournalTitle{Astroparticle
  Physics}, 3, 267 }

\bibitem[{Lesiak-Bzdak {et~al.}(2015)Lesiak-Bzdak, Niederhausen, St\"ossl,
  {et~al.}}]{Aartsen:2015zva:casc}
Lesiak-Bzdak, M., Niederhausen, H., St\"ossl, A., {et~al.} 2015,
  \href{https://inspirehep.net/record/1398539/files/arXiv:1510.05223.pdf}{in
  {Proceedings, 34th International Cosmic Ray Conference (ICRC 2015)}}, 59

\bibitem[{Lundberg {et~al.}(2007)Lundberg, Miocinovic, Burgess, Adams,
  Hundertmark, Desiati, Woschnagg, \& Niessen}]{Lundberg:PHOTONICS}
Lundberg, J., Miocinovic, P., Burgess, T., {et~al.} 2007,
  \href{http://dx.doi.org/10.1016/j.nima.2007.07.143}{\JournalTitle{Nucl.
  Instrum. Meth.}, A581, 619}

\bibitem[{Markov(1960)}]{Markov:1960vja}
Markov, M.~A. 1960,
  \href{http://inspirehep.net/record/1341439/files/C60-08-25-p578.pdf}{in
  {Proceedings, 10th International Conference on High-Energy Physics (ICHEP
  60): Rochester, NY, USA, 25 Aug - 1 Sep 1960}}, 578

\bibitem[{Mohrmann {et~al.}(2015)}]{Aartsen:2015zva:GlobalFit}
Mohrmann, L., {et~al.} 2015,
  \href{https://inspirehep.net/record/1398539/files/arXiv:1510.05223.pdf}{in
  {Proceedings, 34th International Cosmic Ray Conference (ICRC 2015)}}, 21

\bibitem[{Murase {et~al.}(2014)Murase, Inoue, \& Dermer}]{MuraseBlazers:2014}
Murase, K., Inoue, Y., \& Dermer, C.~D. 2014,
  \href{http://dx.doi.org/10.1103/PhysRevD.90.023007}{\JournalTitle{Phys.
  Rev.}, D90, 023007}

\bibitem[{Narsky \& Porter(2013)}]{narsky2013statistical}
Narsky, I., \& Porter, F.~C. 2013, Statistical analysis techniques in particle
  physics: Fits, density estimation and supervised learning (John Wiley \&
  Sons)

\bibitem[{Neronov \& Semikoz(2016)}]{Neronov:2015osa}
Neronov, A., \& Semikoz, D.~V. 2016,
  \href{http://dx.doi.org/10.1016/j.astropartphys.2015.11.002}{\JournalTitle{Astropart.
  Phys.}, 75, 60}

\bibitem[{Nolan {et~al.}(2012)}]{Fermi2FGL}
Nolan, P., {et~al.} 2012,
  \href{http://dx.doi.org/10.1088/0067-0049/199/2/31}{\JournalTitle{Astrophys.
  J. Suppl.}, 199, 31}

\bibitem[{Olive {et~al.}(2014)}]{pdg:Agashe:2014kda}
Olive, K.~A., {et~al.} 2014,
  \href{http://dx.doi.org/10.1088/1674-1137/38/9/090001}{\JournalTitle{Chin.
  Phys.}, C38, 090001}

\bibitem[{Pedregosa {et~al.}(2011)Pedregosa, Varoquaux, Gramfort, Michel,
  Thirion, Grisel, Blondel, Prettenhofer, Weiss, Dubourg,
  {et~al.}}]{pedregosa2011scikit}
Pedregosa, F., Varoquaux, G., Gramfort, A., {et~al.} 2011, \JournalTitle{The
  Journal of Machine Learning Research}, 12, 2825

\bibitem[{{R\"adel}(2016)}]{PhDThesis:Raedel}
{R\"adel}, L. 2016, \JournalTitle{PhD thesis, RWTH Aachen University (in
  preparation)}

\bibitem[{Razzaque(2013)}]{PhysRevD.88.081302}
Razzaque, S. 2013,
  \href{http://dx.doi.org/10.1103/PhysRevD.88.081302}{\JournalTitle{Phys. Rev.
  D}, 88, 081302}

\bibitem[{Reines(1960)}]{Reines:1960}
Reines, F. 1960,
  \href{http://dx.doi.org/10.1146/annurev.ns.10.120160.000245}{\JournalTitle{Ann.
  Rev. Nucl. Part. Sci.}, 10, 1}

\bibitem[{Reines \& Cowan(1956)}]{Reines:1956}
Reines, F., \& Cowan, C.~L. 1956,
  \href{http://dx.doi.org/10.1038/178446a0}{\JournalTitle{Nature}, 178, 446}

\bibitem[{Roberts(1992)}]{Roberts:1992re}
Roberts, A. 1992,
  \href{http://dx.doi.org/10.1103/RevModPhys.64.259}{\JournalTitle{Rev. Mod.
  Phys.}, 64, 259}

\bibitem[{Roulet {et~al.}(2013)Roulet, Sigl, van Vliet, \&
  Mollerach}]{JCAP.2013.01.028}
Roulet, E., Sigl, G., van Vliet, A., \& Mollerach, S. 2013,
  \href{http://stacks.iop.org/1475-7516/2013/i=01/a=028}{\JournalTitle{Journal
  of Cosmology and Astroparticle Physics}, 2013, 028}

\bibitem[{Schoenen(2016)}]{PhDThesis:Schoenen}
Schoenen, S. 2016, \JournalTitle{PhD thesis, RWTH Aachen University (in
  preparation)}

\bibitem[{Schoenen \& Raedel(2015)}]{schoenen2015detection}
Schoenen, S., \& Raedel, L. 2015,
  \href{http://www.astronomerstelegram.org/?read=7856}{\JournalTitle{The
  Astronomer's Telegram}, 7856, 1}

\bibitem[{Schukraft(2013)}]{AnnePhD}
Schukraft, A. 2013, \JournalTitle{PhD thesis, RWTH Aachen University}

\bibitem[{Senno {et~al.}(2016)Senno, Murase, \& Meszaros}]{Senno:2015}
Senno, N., Murase, K., \& Meszaros, P. 2016,
  \href{http://dx.doi.org/10.1103/PhysRevD.93.083003}{\JournalTitle{Phys.
  Rev.}, D93, 083003}

\bibitem[{Stecker(2013)}]{PhysRevD.88.047301}
Stecker, F.~W. 2013,
  \href{http://dx.doi.org/10.1103/PhysRevD.88.047301}{\JournalTitle{Phys. Rev.
  D}, 88, 047301}

\bibitem[{Tamborra {et~al.}(2014)Tamborra, Ando, \&
  Murase}]{Tamborra:2014:Starforming}
Tamborra, I., Ando, S., \& Murase, K. 2014,
  \href{http://dx.doi.org/10.1088/1475-7516/2014/09/043}{\JournalTitle{JCAP},
  1409, 043}

\bibitem[{Wakely \& Horan(2007)}]{TeVCat}
Wakely, S.~P., \& Horan, D. 2007,
  \href{http://indico.nucleares.unam.mx/contributionDisplay.py?contribId=378&confId=4}{in
  {Proceedings, 30th International Cosmic Ray Conference (ICRC 2007)}, Vol.~3},
  1341

\bibitem[{Waxman(2013)}]{WaxmanUpperBound:2013}
Waxman, E. 2013,
  \href{https://inspirehep.net/record/1266927/files/arXiv:1312.0558.pdf}{in
  {Rencontres du Vietnam: Windows on the Universe Quy Nhon, Binh Dinh, Vietnam,
  August 11-17, 2013}}, 161

\bibitem[{Wilks(1938)}]{WilksTheorem}
Wilks, S.~S. 1938,
  \href{http://dx.doi.org/10.1214/aoms/1177732360}{\JournalTitle{Ann. Math.
  Statist.}, 9, 60}

\end{thebibliography}
\end{document}